\documentclass[12pt,preprint]{aastex}
\usepackage{amsmath}

\begin{document}

\title{Exploring the Variable Sky with LINEAR. II. Halo Structure and Substructure Traced by RR Lyrae Stars to 30 kpc} 

\author{
Branimir Sesar\altaffilmark{\ref{Caltech}},
\v{Z}eljko Ivezi\'{c}\altaffilmark{\ref{Washington}},
J.~Scott Stuart\altaffilmark{\ref{LLMIT}},
Dylan M.~Morgan\altaffilmark{\ref{Washington}},
Andrew C. Becker\altaffilmark{\ref{Washington}},
Sanjib Sharma\altaffilmark{\ref{Sydney}},
Lovro Palaversa\altaffilmark{\ref{Geneve}},
Mario Juri\'{c}\altaffilmark{\ref{Steward},\ref{LSST}},
Przemyslaw Wozniak\altaffilmark{\ref{LANL}},
Hakeem Oluseyi\altaffilmark{\ref{FIT}}
}

\altaffiltext{1}{Division of Physics, Mathematics and Astronomy, Caltech,
                 Pasadena, CA 91125\label{Caltech}}
\altaffiltext{2}{University of Washington, Department of Astronomy, P.O.~Box
                 351580, Seattle, WA 98195-1580\label{Washington}}
\altaffiltext{3}{Lincoln Laboratory, Massachusetts Institute of Technology,
                 244 Wood Street, Lexington, MA 02420-9108\label{LLMIT}}
\altaffiltext{4}{Sydney Institute for Astronomy, School of Physics, University
                 of Sydney, NSW 2006, Australia\label{Sydney}}
\altaffiltext{5}{Observatoire astronomique de l'Universit\'{e} de Gen\`{e}ve,
                 51 chemin des Maillettes,
                 CH-1290 Sauverny, Switzerland\label{Geneve}}
\altaffiltext{6}{Steward Observatory, University of Arizona,
                 Tucson, AZ 85121\label{Steward}}
\altaffiltext{7}{LSST Corporation, 933 North Cherry Avenue,
                 Tucson, AZ 85721\label{LSST}}
\altaffiltext{8}{Los Alamos National Laboratory, 30 Bikini Atoll Rd.,
                 Los Alamos, NM 87545-0001\label{LANL}}
\altaffiltext{9}{Florida Institute of Technology, Melbourne,
                 FL 32901\label{FIT}}

\begin{abstract}
We present a sample of $\sim5,000$ RR Lyrae stars selected from the recalibrated
LINEAR dataset and detected at heliocentric distances between 5 kpc and 30 kpc
over $\sim8,000$ deg$^2$ of sky. The coordinates and light curve properties,
such as period and Oosterhoff type, are made publicly available. We analyze in
detail the light curve properties and Galactic distribution of the subset of
$\sim4,000$ type-$ab$ RR Lyrae stars, including a search for new halo
substructures and the number density distribution as a function of Oosterhoff
type. We find evidence for the Oosterhoff dichotomy among field RR Lyrae stars,
with the ratio of the type II and I subsamples of about 1:4, but with a weaker
separation than for globular cluster stars. The wide sky coverage and depth of
this sample allows unique constraints for the number density distribution of
halo RRab stars as a function of galactocentric distance: it can be described as
an oblate ellipsoid with the axis ratio $q = 0.63$ and with either a single or a
double power law with a power-law index in the range $-2$ to $-3$. Consistent
with previous studies, we find that the Oosterhoff type II subsample has a
steeper number density profile than the Oosterhoff type I subsample. Using a
group-finding algorithm EnLink, we detected seven candidate halo groups, only
one of which is statistically spurious. Three of these groups are near globular
clusters (M53/NGC 5053, M3, M13), and one is near a known halo substructure
(Virgo Stellar Stream); the remaining three groups do not seem to be near any
known halo substructures or globular clusters, and seem to have a higher ratio
of Oosterhoff type II to Oosterhoff type I RRab stars than what is found in the
halo. The extended morphology and the position (outside the tidal radius) of
some of the groups near globular clusters is suggestive of tidal streams
possibly originating from globular clusters. Spectroscopic followup of detected
halo groups is encouraged.
\end{abstract}


\section{Introduction}\label{introduction}

Studies of the Galactic halo provide unique insights into the formation history 
of the Milky Way, and for the galaxy formation process in general \citep{hel08}.
One of the main reasons for this uniqueness is that dynamical timescales are
much longer than for disk stars and thus the ``memory of past events lasts
longer'' \citep[e.g.,][]{jhb96, may02}. For example, within the framework of
hierarchical galaxy formation \citep{fbh02}, the spheroidal component of the
luminous matter should reveal substructures such as tidal tails and streams
\citep{jhb96, helmi99, bkw01, har01}. The amount of substructures and the
distribution of their properties like mass, and radial distance can be used to
place constraints on the accretion history of the Galaxy \citep{joh08, sha11}.
The number of these substructures, created due to mergers and accretion over the
Galaxy's lifetime, may provide a crucial test for proposed solutions to the
``missing satellite'' problem \citep{bkw01}. Substructures are expected to be
ubiquitous in the outer halo (galactocentric distance $>15-20$ kpc), and indeed
many have been discovered (for a recent review, see \citealt{ibj12}).
Understanding the number density distribution of stars (i.e., the structure) in
the halo is equally important because its shape and profile affect estimates of
the degree of velocity anisotropy and estimates of the mass of the Milky Way
\citep{dea12, kaf12}. Various luminous tracers, such as main-sequence turn-off
stars, RR Lyrae variables, blue horizontal branch stars, and red giants, are
used to map halo structure and substructures; among them, RR Lyrae stars have
proven to be especially useful.

RR Lyrae stars represent a fair sample of the old halo population \citep{smi04}.
They are nearly standard candles, are sufficiently bright to be detected at
large distances, and are sufficiently numerous to trace the halo substructures
with good spatial resolution \citep{ses10a}. Fairly complete and relatively
clean samples of RR Lyrae stars can be selected using single-epoch colors
\citep{ive05}, and if multi-epoch data exist, using variability
(\citealt{ive00}; QUEST, \citealt{viv01}; \citealt{ses07}; \citealt{dle08}; 
SEKBO, \citealt{kel08}; LONEOS-I, \citealt{mic08}). A useful comparison of
recent RR Lyrae surveys in terms of their sky coverage, distance limits, and
sample size is presented by \citet[see their Table 1]{kel08}.

As an example of the utility of RR Lyrae samples, the period and amplitude of
their light curves may hold clues about the formation history of the Galactic
halo \citep{cat09}. The distribution of RR Lyrae stars in globular clusters in
the period-amplitude diagram displays a dichotomy, first noted by \citet{oos39}.
According to \citet{cat09}, if the Galactic halo was entirely built from smaller
``protogalactic fragments'' like the present-day Milky Way dwarf spheroidal
(dSph) satellite galaxies, the halo should not display this so-called Oosterhoff
dichotomy (see Section~\ref{Oosterhoff} for details) because the dSph galaxies
and their globular clusters are predominantly intermediate between the two
Oosterhoff classes. Whether the present-day halo displays the Oosterhoff
dichotomy is still a matter of contention. Some studies claim detection of
distinct Oo I and and Oo II components \citep{mic08, dle08, spp09}, while others
do not see a clear Oo II component \citep{kin06}.

In order to determine the Oosterhoff class for an RR Lyrae star, a well-sampled
light curve is needed. Most of the above studies used RR Lyrae stars selected
from surveys that were either deep with small sky coverage (e.g., 300 deg$^2$
large SDSS Stripe 82; \citealt{dle08}; \citealt{wat09}; \citealt{ses10a}), or
shallow with wide sky coverage (e.g., ASAS; \citealt{spp09}). LINEAR is a
wide-area survey that provides both depth and a large area\footnote{While this
paper was in preparation, the first analysis of $\sim$12,000 RR Lyrae stars
selected from half of the sky monitored by the Catalina Survey was reported by
\citet{dra13}. For a comparison of their results and our work, see
Section~\ref{sec:disc}.}; RR Lyrae stars from LINEAR are detected to the edge of
the inner halo ($\sim30$ kpc) over a sky area of $\sim8000$ deg$^{2}$. The main 
goals of this paper are to i) present a sample of $\sim$5,000 RR Lyrae stars
selected from the LINEAR database, and ii) quantify their spatial distribution
and the differences, if any, between the behavior of the two Oosterhoff classes.

This paper is the second one in a series based on light curve data collected by
the asteroid LINEAR survey. In the first paper, \citet{ses11} described the
LINEAR survey and photometric recalibration based on SDSS stars acting as a
dense grid of standard stars. In the overlapping $\sim$10,000 deg$^2$ of sky
between LINEAR and SDSS, Sesar et al.~obtained photometric errors of 0.03 mag
for sources not limited by photon statistics, with errors rising to 0.2 mag at
$r\sim18$. LINEAR data provide time domain information for the brightest 4
magnitudes of SDSS survey, with 250 unfiltered photometric observations per
object on average (rising to $\sim$500 along the Ecliptic). Public access to the
recalibrated LINEAR data, including over 5 billion photometric measurements for
about 25 million objects (about three quarters are stars) is provided through
the SkyDOT Web site (\url{https://astroweb.lanl.gov/lineardb/}). Positional
matches to SDSS and 2MASS \citep{skr06} catalog entries are also available for
the entire sample. 

The selection criteria for RR Lyrae stars and analysis of the contamination and
completeness of the resulting sample are described in Section~\ref{selection}.
Estimation of the light curve parameters and distance determination are
discussed in Section~\ref{catalog}, and the period-amplitude distribution in
Section~\ref{Oosterhoff}. The spatial distribution of the resulting samples
is quantified in Section~\ref{number_density}, and the search for halo
substructures is presented in Section~\ref{substructure}. Our results are
discussed and summarized in Section~\ref{sec:disc}. 

\section{Selection of RR Lyrae stars}\label{selection}

In this Section we describe the method used to select RR Lyrae stars from the
recalibrated LINEAR dataset. The selection method is fine-tuned using a training
set of known RR Lyrae stars selected by \citet[hereafter Ses10]{ses10a} from the
SDSS Stripe 82 region. Even though this training set is estimated to be
essentially complete ($\sim99$\%; \citealt{suv12}) and contamination-free, we
confirm these estimates in Sections~\ref{ses10_contamination}
and~\ref{contamination}. In the context of this work, the sample completeness is
defined as the fraction of RR Lyrae stars recovered as a function of magnitude,
and the contamination is defined as the fraction of non-RR Lyrae stars in a
sample.

We start initial selection by selecting point-like (SDSS objtype$=$6) objects
from the LINEAR database that:
\begin{itemize}
\item are located in the region of the sky defined by
$309\arcdeg < {\rm R.A.} < 60\arcdeg$ and $|{\rm Dec}|<1.23\arcdeg$, where both 
SDSS Stripe 82 and LINEAR have uniform coverage,
\item have light curves with at least 15 good observations in LINEAR
(${\rm nPtsGood}\geq15$), and
\item have single-epoch SDSS colors (corrected for extinction using the
\citealt{SFD98} dust map) in these ranges:
\begin{eqnarray}
0.75 < u-g < 1.45 \label{single_epoch_colors1} \\
-0.25 < g-r < 0.4 \\
-0.2 < r-i < 0.2 \\
-0.3 < i-z < 0.3\label{single_epoch_colors2}.
\end{eqnarray}
\end{itemize}
The last criterion limits the acceptable range of single-epoch SDSS colors that
a candidate RR Lyrae star may have \citep{ses10a}. Using a sample of $\sim500$
RR Lyrae stars from the SDSS Stripe 82 region, Ses10 have shown that
Equations~\ref{single_epoch_colors1} to \ref{single_epoch_colors2} encompass the
full range of SDSS colors that a RR Lyrae star may have, irrespective of the
phase. Therefore, by considering LINEAR objects with these single-epoch SDSS
colors we eliminate most non-RR Lyrae stars, and still select all true RR Lyrae
stars. The last criterion reduces the sample of candidates by a factor of eight
to 90,897 candidates.

\subsection{Light curve analysis} 

In the next step, we use an implementation of the {\em Supersmoother} algorithm
\citep{fri84,rei94} to find five most likely periods of variability for the
90,897 LINEAR objects that pass the above cuts. For 22,117 candidate RR Lyrae
stars, Supersmoother returns one or more periods in the $0.2-0.9$ day range
(typical of RR Lyrae stars, \citealt{smi04}); the curves are phased
(period-folded) with each period and a set of SDSS $r$-band templates from Ses10
are fitted to phased data.

Even though LINEAR cameras observe without a spectral filter, the choice of SDSS
$r$-band templates for light curve fitting is an appropriate one. As shown in
Figure~4 from \citet{ses11}, the color term between the LINEAR magnitude and
SDSS $r$-band magnitude is essentially independent of color for blue stars such 
as RR Lyrae stars ($\sim0.02$ mag within $0<g-i<0.5$). This means that the
shapes of RR Lyrae light curves in the LINEAR and SDSS $r$-band will be
identical for all practical purposes (especially so given the systematic error
in LINEAR magnitudes of $\sim$0.03 mag and rapidly increasing photometric
uncertainty at magnitudes fainter than 15 mag, see Figure~12 in
\citealt{ses11}).

The light-curve fitting to estimate the best period and template is performed by
minimizing the robust goodness-of-fit cost function defined in
Equation~\ref{abs_sig_dev} in the least-square sense, with the heliocentric
Julian date (HJD) of peak brightness $HJD_0$, peak-to-peak amplitude $A$, and
peak brightness $m_0$ as free parameters. The quality of a template fit is
defined with a $\chi^2$-like parameter ($L_1$ norm)
\begin{equation}
\zeta = {\rm median}(|m_{\rm observed}^i - m_{\rm template}|/\epsilon_{\rm observed}^i)\label{abs_sig_dev},
\end{equation}
where $m_{observed}$ and $\epsilon_{observed}$ are the observed magnitude and
its uncertainty, $m_{template}$ is the magnitude predicted by the template, and
$i=1, N_{obs}$, where $N_{obs}$ is the number of observations. Here we use the
median to minimize the bias in $\zeta$ due to rare observations with anomalous
(non-Gaussian) errors (e.g., due to image artifacts, cosmic rays). The template
with the lowest $\zeta$ value is selected as the best fit, and the best-fit
parameters are stored.

In addition to these parameters, we also estimate the shape of the folded light
curve using the skewness of the distribution of the medians of magnitudes binned
in phase bins (binned 0.1 in phase). We find this skewness (hereafter $\gamma$) 
to be more robust than the skewness calculated using all data points (not binned
in phase), because it reduces the impact of uneven sampling and filters out
observations that may have unreliable errors (e.g., due to image artifacts,
cosmic rays).

\subsection{Optimization of the selection criteria for RR Lyrae stars}

At the end of the template fitting step, each light curve is characterized with
the following parameters: $\chi^2$ per degree of freedom $\chi^2_{pdf}$, number
of good LINEAR observations $nPtsGood$, period of variability $P$, peak-to-peak
amplitude $A$, peak brightness $m_0$, and light-curve skewness $\gamma$. The
next step is to find the right combination of cuts on these parameters that
yields a sample with as high as possible completeness and as low as possible
contamination for {\em both} type $ab$ and type $c$ RR Lyrae stars (i.e., the
selection is not optimized for either type). The virtually complete and
contamination-free sample of RR Lyrae stars selected by Ses10 greatly simplifies
this process.

For a given trial set of cuts, we tag LINEAR candidates that pass these cuts as
RR Lyrae stars, while those that do not pass cuts are tagged as non-RR Lyrae
stars. The tagged candidates are then positionally matched to the SDSS Stripe 82
sample of RR Lyrae stars to confirm whether the tagging was correct or not.

We have found that the following cuts offer the best trade-off between
contamination and completeness (2\% contamination and 80\% completeness for
objects brighter than $\sim18$ mag, see Sections~\ref{contamination}
and~\ref{completeness}):
\begin{eqnarray}
\chi^2_{pdf} > 1 \label{selection_cuts1} \\
nPtsGood > 100 \label{selection_cuts2} \\
-0.6 < \log (P/day) < -0.046 \label{selection_cuts3} \\
A > 0.3\, {\rm mag} \label{selection_cuts4} \\
m_0 < 17.8\, {\rm mag} \label{selection_cuts5} \\
-1 < \gamma < 0.2\label{selection_cuts6}.
\end{eqnarray}
The $\chi^2_{pdf} > 1$, $nPtsGood > 100$, and $m_0 < 17.8$ mag cuts were
motivated by properties of the LINEAR data set (e.g., the LINEAR faint limit is
at $\sim18$ mag and the median number of non-flagged observations per object is
$\sim200$; \citealt{ses11}), while the cuts on amplitude, period, and skewness
were motivated by light curve properties of RR Lyrae stars (e.g., see Figure 11
in \citealt{ses07} and Figure 16 in \citealt{ses11}). In total, the above
criteria tag 226 objects from SDSS Stripe 82 that also have LINEAR light curves
as RR Lyrae stars. 

\subsection{Contamination in the \citet{ses10a} sample of RR Lyrae stars}\label{ses10_contamination}

In Sections~\ref{contamination} and~\ref{completeness}, the Ses10 sample of RR
Lyrae stars is used as the ``ground truth'' when estimating the efficiency of
the above selection algorithm. Before proceeding further, it seems prudent to
verify the level of contamination in this ``ground truth'' sample using more
numerous observations provided by the LINEAR data set.

The key factor that influences the classification of an object is its period. If
the period is incorrect, a true RR Lyrae star may be rejected or a non-RR Lyrae
stars may be accepted. Thus, a good starting point for finding possible
contaminants is to search for objects that have different periods when derived
from different light curve data sets.

To check for contamination by non-RR Lyrae stars in the Ses10 sample of RR Lyrae
stars, we compare two sets of phased LINEAR light curves of Ses10 RR Lyrae
stars. The first set is phased with periods derived from LINEAR data, and the
second set is phased with periods derived from SDSS Stripe 82 data. For most
bright ($m_0 < 17$) and well-sampled LINEAR objects, the LINEAR and SDSS periods
agree within a root-mean-square scatter (rms) of 0.3 sec. However, there are two
Stripe 82 objects (RR Lyrae ID 747380 and 1928523 from Ses10) for which the
LINEAR period provides a much smoother phased light curve than the period
derived from SDSS Stripe 82 data. These LINEAR periods are much shorter than the
SDSS periods ($\sim$0.28 days vs.~$\sim$0.6 days), and challenge the initial RR
Lyrae classification of the two objects.

\begin{figure}
\epsscale{1.0}
\plotone{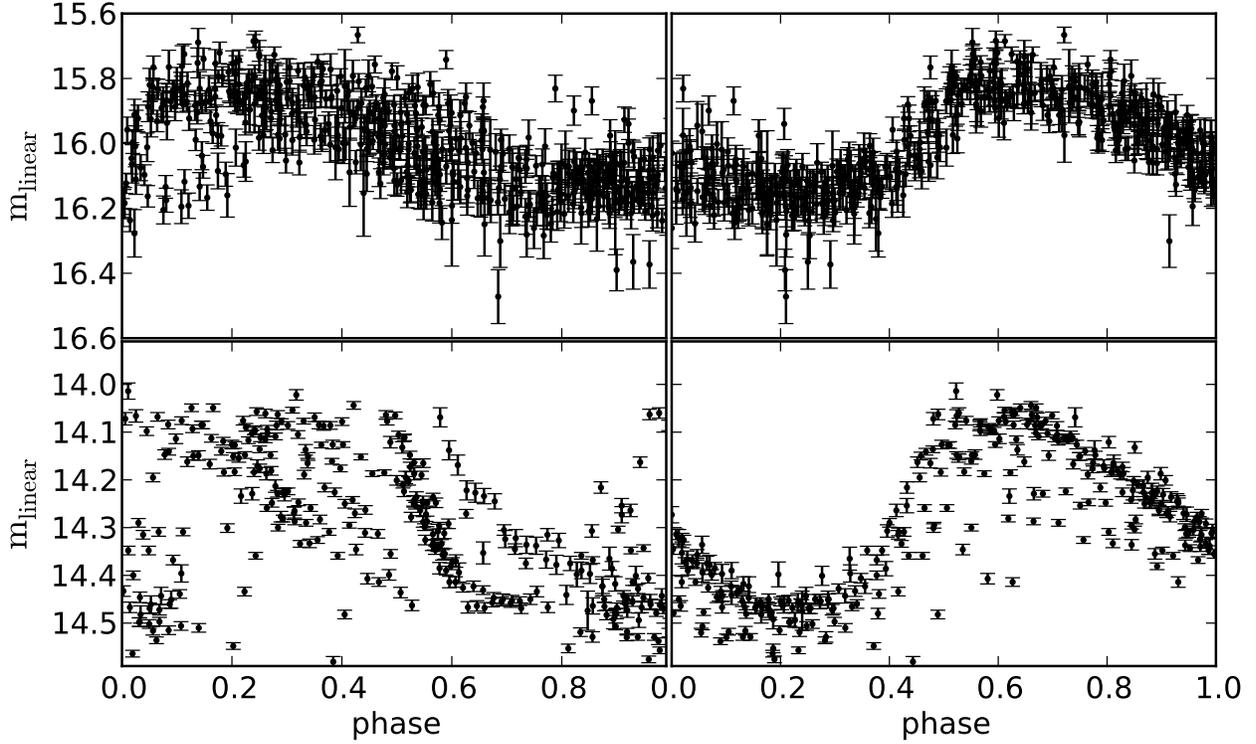}
\caption{
A comparison of LINEAR light curves phased using periods derived from SDSS
({\em left}) and LINEAR data ({\em right}). The LINEAR light curves for
\citet{ses10a} RR Lyrae with IDs 747380 and 1928523 are shown in top and bottom
panels, respectively. For these two stars, the LINEAR periods seem to be more
accurate since the light curves phased using LINEAR periods, shown on the right,
are smoother than light curves phased using SDSS periods. The LINEAR periods are
much shorter than SDSS periods ($\sim$0.28 days vs.~$\sim$0.6 days) indicating
that the two stars are more likely to be type $c$ RR Lyrae stars and not type
$ab$ RR Lyrae stars as originally classified by \citet{ses10a}.
\label{sdss_vs_linear_lc}}
\end{figure}

After a visual inspection of their phased light curves, shown in
Figure~\ref{sdss_vs_linear_lc}, we conclude that these objects are likely type
$c$ RR Lyrae stars, instead of type $ab$ RR Lyrae stars as originally classified
by Ses10. Even though the classification changed from one RR Lyrae type to
another, these stars are still RR Lyrae stars and should not be considered as
contaminants in the Ses10 sample. Based on this analysis, we confirm our initial
assumption that the Ses10 sample is essentially free of contamination.

\subsection{Contamination in the LINEAR sample of RR Lyrae stars}\label{contamination}

The contamination, or the fraction of non-RR Lyrae stars in a sample of
candidate RR Lyrae stars, is an important quantity that needs to be known (and
minimized) before the Galactic halo is mapped. As an illustration of the impact
of contamination on Galactic halo number density maps, consider RR Lyrae samples
obtained by \citet{ses07} and Ses10. Due to the smaller number of epochs
available at the time, the \citet{ses07} sample of RR Lyrae stars had a higher
contamination than a more recent sample constructed by Ses10 ($\sim30\%$
vs.~close to zero contamination in Ses10). The result of contamination in the
\citet{ses07} sample was the appearance of false halo overdensities in their
halo number density maps (e.g., overdensities labeled D, F, H, I, K, and L in
Figure 13 by \citealt{ses07}), which are not present in density maps obtained
using a much cleaner Ses10 sample (see Figure 11 by Ses10). These false
overdensities observed by \citet{ses07} consist of variable, non-RR Lyrae stars
(mainly $\delta$ Scuti stars), which were projected in distance far into the
halo due to the incorrect assignment of absolute magnitudes (i.e., $M_V = 0.6$
mag typical of RR Lyrae stars was assigned when the true absolute magnitude
value is much lower). 

\begin{figure}
\epsscale{1.0}
\plotone{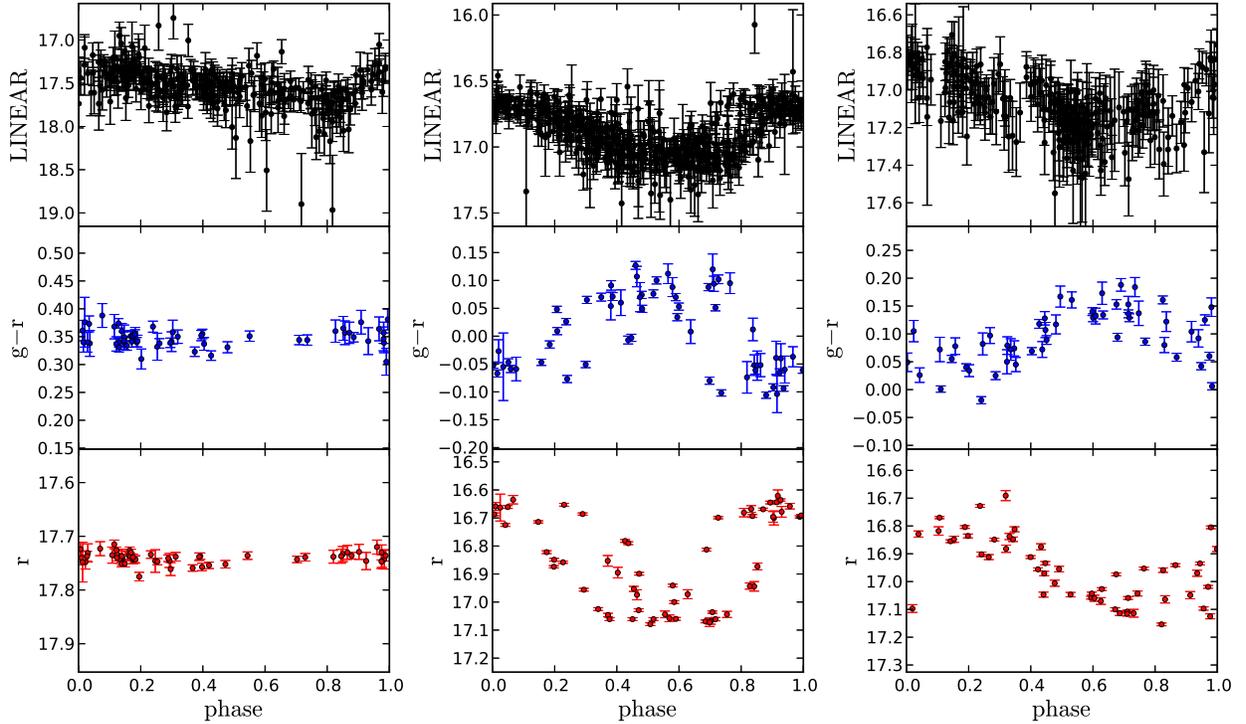}
\caption{
The LINEAR and SDSS $g-r$ and $r$-band light curves of some objects tagged as RR
Lyrae stars based on LINEAR data, but not tagged as RR Lyrae stars by
\citet{ses10a}. The object on the left is clearly not a RR Lyrae star, and it
was most likely falsely tagged as a RR Lyrae star due to its noisy LINEAR data.
There are three more objects with noisy LINEAR light curves that were tagged as
RR Lyrae stars, but they are not shown in this plot. The object in the middle is
possibly a Bla\v{z}ko or a double-mode (type $d$) RR Lyrae star, and the object 
on the right is probably a variable non-RR Lyrae star.
\label{linear_vs_s82}}
\end{figure}

Out of 226 LINEAR objects tagged as RR Lyrae stars in the SDSS Stripe 82 by our
selection algorithm, only 6 (or $\sim3\%$) are not in the Ses10 Stripe 82 sample
of RR Lyrae stars. One possibility is that these objects are non-RR Lyrae stars.
Alternatively, some or all of them may be true RR Lyrae stars that were
overlooked by Ses10 and therefore were not included in their final sample (i.e.,
the Ses10 sample may not be complete). To find whether any of these 6 stars are
RR Lyrae stars, we phase their LINEAR and SDSS $g$- and $r$-band observations
using the best-fit period determined from LINEAR data, and plot their phased
light curves in Figure~\ref{linear_vs_s82} for visual inspection.

Their phased light curves reveal that 4 out 6 objects have noisy LINEAR light
curves (objects are faint and have $m_0 > 17$ mag), and show no significant
variability in SDSS data. Noisy LINEAR light curves are the most likely reason
why these spurious, non-variable objects end up in our RR Lyrae sample. The
remaining two objects show significant variability in SDSS data: one is possibly
a Bla\v{z}ko or a double-mode (type-$d$) RR Lyrae star while the other variable
object is probably not a RR Lyrae star.

The above analysis suggests that our selection algorithm produces a RR Lyrae
sample where only up to $\sim$2\% of objects are non-RR Lyrae stars. The
majority of contaminants are spurious, non-variable objects with noisy LINEAR
data. Since the RR Lyrae sample is mostly contaminated at the faint end, special
attention needs to be given to distant halo overdensities as these are more
likely to contain non-RR Lyrae stars and therefore, more likely to be spurious. 
This analysis also suggests that the completeness of the Ses10 sample is very
high, with plausibly only one RR Lyrae star missed by Ses10 in the range
$r < 17$ (the magnitude range probed by LINEAR).

\subsection{Completeness of the LINEAR sample of RR Lyrae stars}\label{completeness}

The completeness, or the fraction of RR Lyrae stars recovered as a function of
magnitude, is another important quantity that needs to be understood before the
spatial distribution of RR Lyrae stars can be analyzed. To quantify
completeness, we again use the Ses10 sample of RR Lyrae stars as the ``ground
truth'' and assume the sample is complete and clean based on the analyses
presented in the previous two subsections.

The completeness as a function of peak magnitude $m_0$ is defined as the ratio
\begin{equation}
f_{\rm completeness}(m_0)=N_{\rm selected}(m_0)/N_{\rm all}(m_0),
\end{equation}
where $N_{\rm selected}$ is the number of SDSS Stripe 82 RR Lyrae stars that
have been tagged by our selection algorithm and $N_{\rm all}$ is the number of
all SDSS Stripe 82 RR Lyrae stars in a magnitude bin centered on $m_0$. The peak
brightness of an SDSS Stripe 82 RR Lyrae star in the LINEAR photometric system
($m_0$) is calculated using its best-fit peak brightness in the SDSS $r$-band
light curve ($r_0$; see Table 2 in \citealt{ses10a}) as
\begin{equation}
m_0 = r_0 + 0.0574,\label{m_0}
\end{equation}
where the 0.0574 mag offset accounts for a small magnitude zero-point shift
between SDSS and LINEAR photometric systems (see Equation 6 in \citealt{ses11}).
A comparison of synthetic and observed $m_0$ values shows that the two are
similar within 0.04 mag, as estimated by their rms scatter.

The two types of RR Lyrae stars show a different dependence of completeness on
peak magnitude $m_0$, as illustrated in Fig.~\ref{completeness_plot}. While the
completeness of type $ab$ RR Lyrae stars is estimated at $\sim80\%$ and is
seemingly independent of magnitude for $m_0 < 17.2$ (corresponding to
heliocentric distances from 5 kpc to 23 kpc), the completeness of type $c$ RR
Lyrae shows a strong dependence on magnitude, with low completeness at the
bright and faint ends, and a peak at $m_0\sim16.2$ (but note the size of Poisson
error bars).

\begin{figure}
\epsscale{1.0}
\plotone{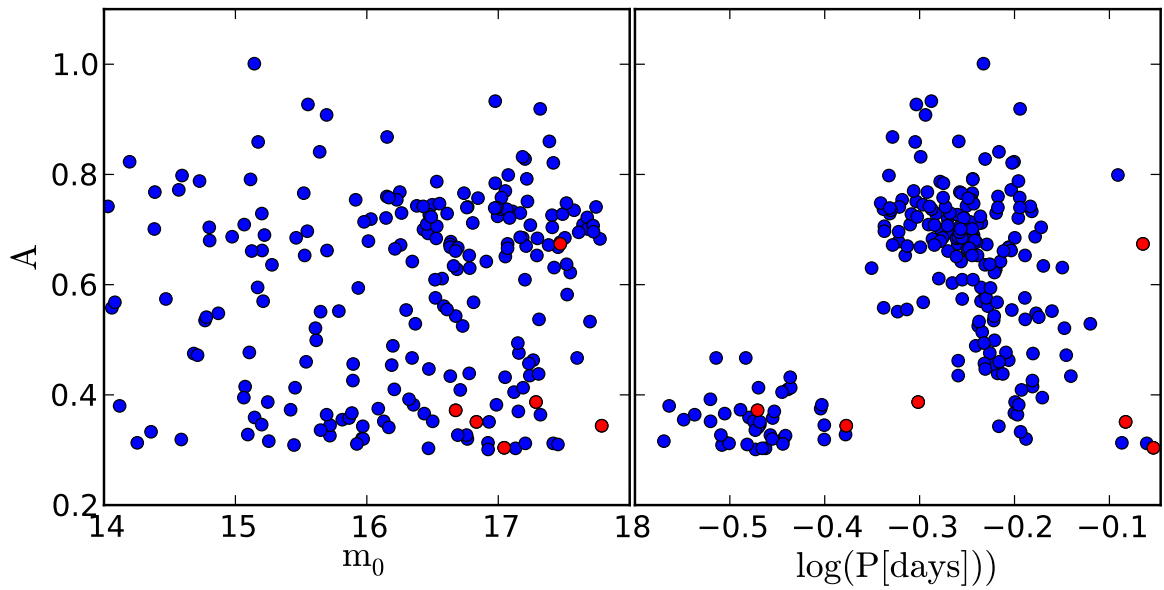}
\caption{
The distribution of LINEAR objects tagged as RR Lyrae stars in amplitude
vs.~peak brightness ({\em left}) and amplitude vs.~period diagrams
({\em right}). LINEAR objects present in the \citet{ses10a} sample of RR Lyrae
stars are shown as blue dots, while those not found in the \citet{ses10a} sample
are shown as red dots. The LINEAR and SDSS Stripe 82 light curves of objects
shown as red dots are compared in Figure~\ref{linear_vs_s82}.
\label{amp_logP_m0}}
\end{figure}

\begin{figure}
\epsscale{1.0}
\plotone{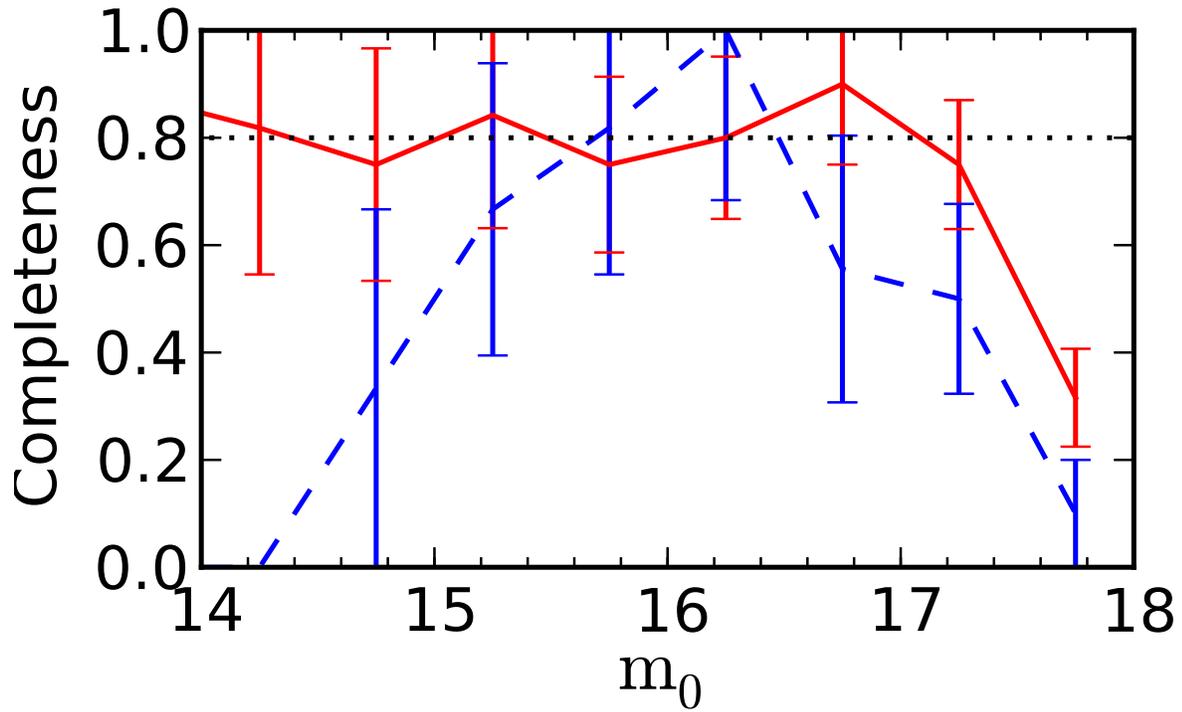}
\caption{
The completeness, or the fraction of RR Lyrae stars recovered as a function of
magnitude for type $ab$ ({\em solid}) and type $c$ ({\em dashed}) RR Lyrae stars
selected in LINEAR. The sample of type $ab$ RR Lyrae stars is about 80\%
complete between 5 and 23 kpc ($14  < m_0 < 17.2$), but do note the Poisson
errorbars.
\label{completeness_plot}}
\end{figure}

Detailed analysis has demonstrated that the amplitude cut $A>0.3$ mag, together
with low typical amplitudes of RRc stars ($A\sim0.3$ mag, see
Figure~\ref{amp_logP_m0}) are the main reasons for low completeness of RRc stars
(the increase in completeness of RRc stars towards $m_0\sim16.2$ is likely due
to Poisson noise). We could have lowered the cut on amplitude to include more
RRc stars, but that would then increase the overall contamination of the more
numerous RRab sample, which we wanted to keep low per discussion in
Section~\ref{contamination}.

\section{LINEAR catalog of RR Lyrae stars}\label{catalog}

There are 533,189 point-like (SDSS objtype$=$6) objects in the LINEAR database
that satisfy conditions given by
Equations~\ref{single_epoch_colors1}--\ref{single_epoch_colors2}
and~\ref{selection_cuts1}--\ref{selection_cuts2}. Out of this sample, we have
selected 4067 type $ab$ (hereafter RRab stars) and 834 type $c$ RR Lyrae stars
(hereafter, RRc stars) following the procedure described in
Section~\ref{selection}. Equatorial J2000.0 right ascension and declination of
selected RRab and RRc stars are listed\footnote{Note that this catalog does not 
contain RR Lyrae stars that are located in SDSS Stripe 82 region.} in
Table~\ref{table1}. For a catalog of RR Lyrae stars in SDSS Stripe 82, we
instead suggest the more complete and deeper Ses10 catalog be used.

\subsection{Final estimation of light curve parameters}\label{final_estimation}

\begin{figure}
\epsscale{0.8}
\plotone{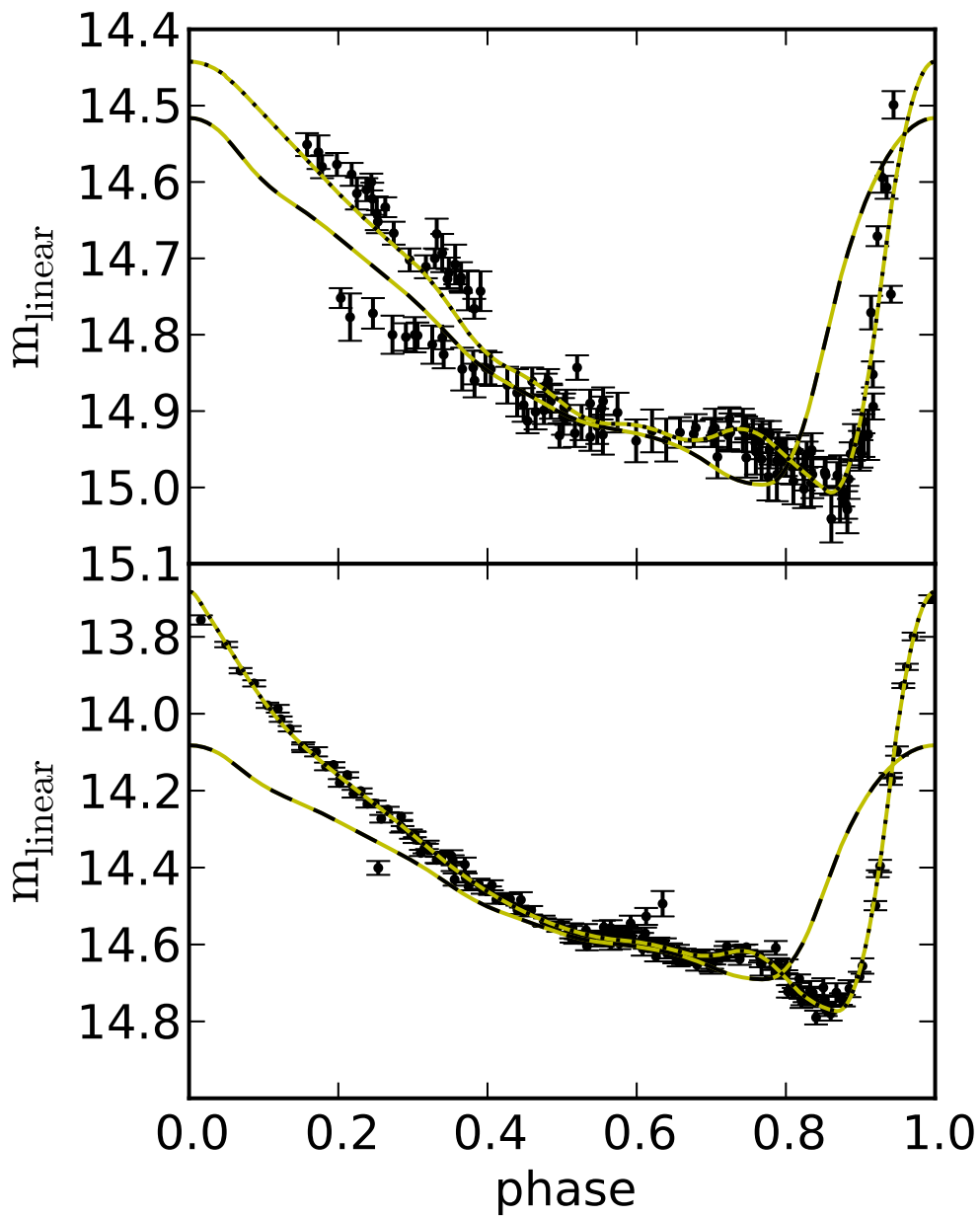}
\caption{
{\em Top}: A phased light curve of a Bla\v{z}ko-variable RRab star. The dashed
line shows the best-fit template from Ses10, while the dotted line shows the
best-fit template created from the LINEAR RRab light curve set. {\em Bottom}: An
example of a RRab light curve where the best-fit template from Ses10
({\em dashed line}) underestimates the amplitude and does not adequately model
the observed data. Note how the best-fit template created from the LINEAR RRab
light curve set ({\em dotted line}) provides a much better fit.
\label{underestimated_amp}}
\end{figure}

Visual inspection of phased light curves has revealed that a non-negligible
number of LINEAR RR Lyrae stars have underestimated best-fit light curve
amplitudes. As shown in the top plot of Figure~\ref{underestimated_amp}, some of
these stars have two or more different maxima and are most likely undergoing
light curve modulations (i.e., the Bla\v{z}ko effect; \citealt{bla07, bk11}).
Determining the true amplitude of such stars may not even be possible as the
Bla\v{z}ko cycle does not always repeat regularly \citep{cha10, kol11, sod11}.

In other cases (bottom plot in Figure~\ref{underestimated_amp}), the light curve
amplitude is underestimated because the best-fit template does not adequately
model the observed light curve. That some light curves are not adequately
modeled is expected because the light curve template set provided by Ses10 is
not all-inclusive (see Figure 7 in Ses10). This inadequate modeling does not
affect the selection procedure as the quality of a template fit ($\zeta$
parameter; see Equation~\ref{abs_sig_dev}) is not used during selection. The
amplitudes, which are used during selection, are underestimated only for RRab
stars with light curve amplitudes greater than 0.5 mag, and such stars are
already above the $A>0.3$ mag selection cut (Equation~\ref{selection_cuts4}).
However, inadequate light curve modeling is a problem as amplitudes (used in
Section~\ref{Oosterhoff}) and flux-averaged magnitudes (used in
Section~\ref{distances}) are derived from best-fit model light curves.

We address the issue of inadequate template fits by re-fitting LINEAR RR Lyrae
light curves with new templates created from the LINEAR RR Lyrae light curve set
itself, following procedure from Ses10. The new templates are constructed by
interpolating a B-spline through phased light curves of $\sim400$ brightest,
well-sampled, and visually inspected LINEAR RRab and RRc light curves that do
not seem to be affected by Bla\v{z}ko variations. These templates are normalized
to the ${\rm [0,1]}$ range in magnitude and then fit to all LINEAR RR Lyrae
light curves. The final light curve parameters are listed in Table~\ref{table1}.

We emphasize that the point of constructing these templates is simply to provide
more accurate model light curves for LINEAR RR Lyrae stars, as these model light
curves are used later in the paper. We did not attempt to prune the template set
by averaging templates with similar shapes (as done by Ses10), and do not
suggest that this new template set should replace the light curve template set
constructed by Ses10. However, we do provide the new templates to support future
work at extending RR Lyrae template light curves (templates are provided as
supplementary data in the electronic edition of the journal).

Table~\ref{table1} also contains 447 RRab and 336 RRc stars from the LINEAR
Catalog of Variable Stars (Palaversa, L.~et al., submitted to AJ). These RR
Lyrae stars were missed by our selection algorithm and are included for
completeness. However, they are not used in the analysis below and their
exclusion does not significantly change our results.

\subsection{Heliocentric Distances}\label{distances}

The heliocentric distances of RR Lyrae stars, $D$, are calculated as
\begin{equation}
D = 10^{(\langle m \rangle - M_{RR})/5+1}/1000\, {\rm kpc},
\end{equation}
where $\langle m \rangle$ is the flux-averaged LINEAR magnitude and $M_{RR}$ is
the absolute magnitude of RR Lyrae stars in the LINEAR bandpass. The
flux-averaged magnitude is calculated by first converting the best-fit model
light curve, $A\, T(\phi)+m_0$, into flux units ($A$, $T(\phi)$, and $m_0$ are
the best-fit amplitude, template, and peak brightness, respectively). This curve
is then integrated and the result is converted back to magnitudes. The
flux-averaged magnitudes, listed in Table~\ref{table1}, are also corrected for
interstellar medium (ISM) extinction
$\langle m \rangle = \langle m \rangle_{not\, corrected} - {\rm rExt}$, where
${\rm rExt} = 2.751E(B-V)$ is the extinction in SDSS $r$ band, and $E(B-V)$
color excess is provided by the \citet{SFD98} dust map.

For RRab stars, we adopt $M_{RR} = 0.6\pm0.1$ as their absolute magnitude. The
absolute magnitude was calculated using the \citet{chaboyer99}
$M_V-{\rm [Fe/H]}$ relation
\begin{equation}
M_V = (0.23\pm0.04){\rm [Fe/H]} + (0.93\pm0.12)\label{abs_mag},
\end{equation}
where we assume that the metallicity of RRab stars is equal to the median
metallicity of halo stars (${\rm [Fe/H]=-1.5}$; \citealt{ive08}). We also assume
that the absolute magnitudes of RRab stars in the LINEAR and Johnson $V$
bandpasses are approximately equal. The estimate of the uncertainty in absolute
magnitude is detailed in the next paragraph. For RRc stars, we simply adopt
$M_{RR} = 0.5\pm0.1$ mag \citep{kol12}.

There are three significant sources of uncertainty in the adopted absolute
magnitude. First, the $M_{RR} \approx M_V$ approximation is uncertain at
$\sim0.04$ mag level (in rms). This uncertainty was estimated by comparing
LINEAR and $V$-band flux-averaged magnitudes of RR Lyrae stars which have
multi-epoch data from SDSS Stripe 82. The $V$-band flux-averaged magnitudes were
calculated from synthetic $V$-band light curves following Section 4.1 by Ses10.
Second, the metallicity dispersion in the Galactic halo is about
$\sigma_{[Fe/H]}=0.3$ dex \citep{ive08}, and introduces about
$\sigma_{M_V}^{[Fe/H]}=0.07$ mag of uncertainty due to the assumption that all
RRab stars have the same metallicity. And third, there is about
$\sigma_{M_V}^{ev}=0.08$ mag of uncertainty due to RR Lyrae evolution off the
zero-age horizontal branch \citep{vz06}. By adding all these uncertainties in
quadrature, the final uncertainty in the absolute magnitude of RRab stars is
about 0.1 mag, implying $\sim5\%$ fractional uncertainty in distance.

In the rest of this work we only use RRab stars. Type $c$ RR RRLyrae stars are
not used due to their much lower completeness (see Section~\ref{completeness}
and Figure~\ref{completeness_plot}).

\section{Period-Amplitude Distribution}\label{Oosterhoff}

As suggested by \citet{cat09}, the period-amplitude distribution of RR Lyrae
stars may hold clues about the formation history of the Galactic halo. Catelan
points to a sharp division (a dichotomy first noted by \citealt{oos39}) in the
average period of RRab stars in Galactic globular clusters,
$\langle P_{ab} \rangle$; there are Oosterhoff type I (Oo I) globular clusters
with $\langle P_{ab} \rangle \sim0.55$ days, Oosterhoff type II (Oo II) globular
clusters with $\langle P_{ab} \rangle \sim0.65$ days, and very few clusters with
$\langle P_{ab} \rangle$ in between. On the other hand, the dwarf spheroidal
satellite galaxies and their globular clusters fall preferentially on the
``Oosterhoff gap'' ($0.58 < \langle P_{ab} \rangle < 0.62$; see his Figure 5).
\citet{cat09} further argues that, if the Oosterhoff dichotomy is present in the
period-amplitude distribution of {\em field} halo RRab stars, then the Galactic
halo could not have been entirely assembled by the accretion of dwarf galaxies
resembling the present-day Milky Way satellites. In light of these conclusions, 
it is interesting to examine the period-amplitude distribution of LINEAR
RRab stars and see whether the Oosterhoff dichotomy is also present among the
Galactic halo field RRab stars.

A period-amplitude diagram for RRab stars listed in Table~\ref{table1} is
shown in Figure~\ref{period_vs_amp}. A locus of stars is clearly visible in this
diagram. We trace this locus by binning $\log P$ values in narrow amplitude
bins, and then calculate the median $\log P$ value in each bin. To model the
locus in the $\log P$ vs.~amplitude diagram, we fit a quadratic function to
medians and obtain
\begin{equation}
\log P = -0.16625223 -0.07021281A -0.06272357A^2\label{OoI_line}
\end{equation}
as the best fit. The solid line in Figure~\ref{period_vs_amp} ({\em top}) is
very similar to the period-amplitude line of RRab stars in the globular cluster
M3 (see Figure 3 by \citealt{cac05}). Since M3 is the prototype Oo I globular
cluster, we label the main locus as the Oo I locus.

The contours in Figure~\ref{period_vs_amp} ({\em top}) seem to indicate a
presence of a second locus of RRab stars to the right of the Oo I locus. To
examine this in more detail, we calculate the period shift, $\Delta \log P$, of
RRab stars from the Oo I locus and at a fixed amplitude. The distribution of
$\Delta \log P$, shown in Figure~\ref{period_vs_amp} ({\em bottom}), is centered
on zero (the position of the Oo I locus), and has a long-period tail. Even
though we do not see the clearly displaced secondary peak that is usually
associated with an Oo II component (see Figure 21 by \citealt{mic08}), hereafter
we will refer to stars in the long-period tail as Oo II RRab stars.

To separate the two Oosterhoff types, we model the Oo I peak with a Gaussian and
find that this Gaussian roughly ends at $\Delta \log P=0.05$. RRab stars with
$\Delta \log P < 0.05$ (where period is measured in days) are tagged as Oo I
RRab stars, and those with $\Delta \log P\ge0.05$ are tagged as Oo II RRab
stars. We find the ratio of Oo II to Oo I RRab stars in the halo to be 1:4
(0.25). A similar ratio was found by \citet{mic08} and \citet{dra13} (26\% and
24\%, respectively).

\section{Number Density Distribution}\label{number_density}

In this Section, we introduce a method for estimating the number density
distribution of RR Lyrae stars that is less sensitive to the presence of halo
substructures (streams and overdensities). The method is first illustrated and
tested on a mock sample of RR Lyrae stars drawn from a known number density
distribution. The purpose of this test is to estimate the precision of the
method in recovering the input model. At the end of this section, the method is
applied to the observed spatial distribution of LINEAR RRab stars.

We begin by drawing ten mock samples of RRab stars from the following number
density distribution:
\begin{eqnarray}
\rho_{model}(X,Y,Z) = \rho_\sun^{RR}\left(\frac{R_\sun}{r}\right)^n\label{RR_model} \\
r = \sqrt{X^2+Y^2+(Z/q)^2}\label{r_distance},
\end{eqnarray}
where $\rho_\sun^{RR}=4.5$ kpc$^{-3}$ is the number density of RRab stars at the
position of the Sun ($R_\sun=8$ kpc or
($X_\sun$, $Y_\sun$, $Z_\sun$)~$=$~(8, 0, 0) kpc), $q=0.71$ is the ratio of
major axes in the $Z$ and $X$ directions indicating that the halo is oblate
(flattened in the $Z$ direction), and $n=2.62$ is the power-law index. The above
model was motivated by \citet{jur08}, who used a similar model to describe the
number density distribution of halo main-sequence stars selected from SDSS. The
$X$, $Y$, and $Z$ are coordinates in the Cartesian galactocentric coordinate
system
\begin{eqnarray}
X = R_\sun - D\cos l\cos b, \\
Y = -D\sin l\cos b, \\
Z = D \sin b,\label{XYZ_coordinates}
\end{eqnarray}
where $l$ and $b$ are Galactic longitude and latitude in degrees, respectively.

The number density model defined by Equation~\ref{RR_model} is assumed to be a
fair representation of the actual number density distribution of halo RRab
stars. The parameters used in the above model were selected based on previous
studies of the Galactic halo. \citet{sji11} have found that a two parameter, 
single power-law ellipsoid model (i.e., Equation~\ref{RR_model}) with
$q=0.71\pm0.01$ and $n=2.62\pm0.04$ provides a good description of the number
density distribution of halo main-sequence stars within 30 kpc of the Galactic
center. Since halo main-sequence stars are progenitors of RR Lyrae stars, it is 
reasonable to assume that shapes of their number density distributions are
similar as well. The number density of RRab stars at the position of the Sun
($\rho_\sun^{RR}$) has been estimated by several studies so far, yielding values
ranging from 4 to 5 kpc$^{-3}$ \citep{psb91,sun91,vz06}.

The mock samples were generated using
{\em Galfast}\footnote{\url{http://mwscience.net/trac/wiki/galfast}} code, which
provides the position, magnitude and distance modulus for each star. To simulate
the uncertainty in heliocentric distance characteristic of the LINEAR sample of
RRab stars, we add Gaussian noise to true distances provided by Galfast ($D_0$)
using a 0.1 mag wide Gaussian centered at zero,
$D = D_0 10^{\mathcal{N}(0,0.1)/5}$.

The presence of halo substructure is simulated by adding a clump of about 620
stars to each mock sample (i.e., about 17\% of stars are in the substructure)
and by distributing them in a uniform sphere 2 kpc wide and centered on
($X$, $Y$, $Z$)$=$(8, 0, 10) kpc (i.e., roughly in the center of the probed
volume of the Galactic halo). Finally, each sample is trimmed down to match the
spatial coverage of the LINEAR sample of RRab stars using the following cuts:
\begin{eqnarray}
b > 30\arcdeg\label{spatial_cuts1}, \\
\delta_{J2000} < -1.2 \, \alpha_{J2000} + 362\arcdeg, \\
-4\arcdeg  < \delta_{J2000} < 72\arcdeg, \\
\delta_{J2000} < 0.73 \, \alpha_{J2000} - 26.6\arcdeg, \\
\delta_{J2000} > 1.38 \, \alpha_{J2000} - 338.52\arcdeg, \\
5\, {\rm kpc} < D < 23\, {\rm kpc}, \\
Z > 3\, {\rm kpc}\label{spatial_cuts2},
\end{eqnarray}
where $\alpha_{J2000}$ and $\delta_{J2000}$ are equatorial right ascension and
declination in degrees, respectively. The penultimate cut limits the samples to 
distances where the LINEAR sample of RRab stars is 80\% complete, and with the
last cut we minimize possible contamination by thick disk stars. Note that mock
samples {\em do not} suffer from any incompleteness or contamination. The reason
we are applying the above cuts to mock samples is because the same cuts will
later be applied to the observed sample of LINEAR RRab stars, which does suffer
from incompleteness at greater distances and may have some contamination from
thick disk RRab stars.

The spatial distribution of stars in each mock sample traces the underlying
number density distribution. To compute the number density of stars at some
$XYZ$ position in the Galactic halo, we use a Bayesian estimator developed by
\citet[also see Chapter 6.1 of \citealt{ive13}]{ive05}:
\begin{equation}
\rho(X, Y, Z) = \frac{3(m+1)}{4\pi}\frac{1}{\sum_{k=1}^{N_{nn}} d_k^3} = \frac{111}{4\pi}\frac{1}{\sum_{k=1}^{8} \left[(X-X_k)^2 + (Y-Y_k)^2 + (Z-Z_k)^2\right]^{3/2}}\label{simple_bayes},
\end{equation}
where $N_{nn} = 8$ is the number of nearest neighbors to which the distance $d$
in a 3-dimensional space is calculated, $m=N_{nn}(N_{nn}+1)/2 = 36$, and
($X_k$, $Y_k$, $Z_k$) is the position of the $k$-th nearest neighbor in
Cartesian galactocentric coordinates. The volume of the Galactic halo probed
by LINEAR RR Lyrae stars is binned in $0.1\times0.1\times0.1$ kpc$^3$ bins, and
the number density is computed for each bin. The bins outside the volume
specified by Equations~\ref{spatial_cuts1}--\ref{spatial_cuts2} are removed to
minimize edge effects. In total, the number density is calculated for about 6
million bins.

Having calculated number densities on a grid, we can now fit
Equation~\ref{RR_model} to computed number densities in order to find the
best-fit $n$, $q$, and $\rho_\sun^{RR}$. Standard $\chi^2$ minimization
algorithms (e.g., the Levenberg-Marquardt algorithm; \citealt{pre92}) are
susceptible to outliers in data, such as unidentified overdensities, and produce
biased results unless outliers are removed. Since we would rather avoid {\em ad 
hoc} removal of suspected outliers, we use a fitting method that is instead
robust to the presence of outliers.

The basic principle of our fitting method is illustrated in
Figure~\ref{logdRho_hists}. When the correct model is used, the height of the
distribution of $\Delta\log\rho = \log\rho - \log\rho_{model}$ values is greater
than when an incorrect model is used. The influence of substructures is
attenuated because overdense regions have highly positive values of
$\Delta\log\rho$ (i.e., they are in the wings of the distribution), and thus
have little effect on the height of the $\Delta\log\rho$ distribution. Note that
in this method, $\rho_\sun^{RR}$ is not a free parameter; $\rho_\sun^{RR}$ is
simply estimated as the mode of the $\Delta\log\rho$ distribution raised to the 
power of 10.

Due to sparseness of the sample (i.e., Poisson noise), $\log\rho$ values will
have a certain level of uncertainty. This uncertainty, or the average random
error in $\log\rho$, can be estimated from the rms scatter of $\Delta\log\rho$
values for the correct model. We find this value to be $\sim0.2$ dex. While the
average random error can be decreased by increasing the number of nearest
neighbors used when computing the density, the downside is an increase in edge
effects if the sample is too sparse.

The best-fit values of $n$ and $q$ parameters are determined by measuring the
height of the $\Delta\log\rho$ distribution on a fixed $n$ vs.~$q$ grid. A
``height'' map for one of the mock RR Lyrae samples generated using Galfast is
shown in Figure~\ref{peak_maps} ({\em left}). Due to sparseness of the sample
(Poisson noise), the best-fit values of $n$ and $q$ parameters will not
necessarily be exactly the same as input $n$ and $q$ values. To estimate the
{\em statistical} uncertainty in best-fit $n$ and $q$ parameters due to Poisson 
noise, we do the fitting on all ten mock samples and analyze the distribution of
best-fit parameters. We find that $\rho_\sun^{RR}=4.4\pm0.6$ kpc$^{-3}$,
$q=0.73\pm0.05$, and $n=2.63\pm0.13$, where the errors represent the rms scatter
(recall that the true values are $\rho_\sun^{RR}=4.5$ kpc$^{-3}$, $q=0.71$, and
$n=2.62$).

Finally, we apply the above procedure to the observed sample of LINEAR RRab
stars. We find that the distribution of RRab stars in the Galactic halo between 
5 and 23 kpc can be modeled as an oblate, single power-law ellipsoid
(Equation~\ref{RR_model}) with the oblateness $q=0.63\pm0.05$ and power-law
index $n=2.42\pm0.13$. The uncertainties on these parameters were adopted from
the analysis of mock samples described above. The best-fit values are consistent
within the 95\% confidence limit with values determined by \citet{sji11} for the
number density distribution of halo main-sequence stars ($q=0.71\pm0.01$,
$n=2.62\pm0.04$). The number density of RRab stars at the position of the Sun is
$\rho_\sun^{RR}=4.5\pm0.6$ kpc$^{-3}$, or $\rho_\sun^{RR}=5.6\pm0.8$ kpc$^{-3}$
once the measured number density is increased by 20\% to account for
completeness of the LINEAR sample of RRab stars (estimated at 80\% in
Section~\ref{completeness}). Adopting $q=0.71$ and $n=2.62$ from \citet{sji11}, 
we obtain $\rho_\sun^{RR}=4.9\pm0.6$ kpc$^{-3}$, or $\rho_\sun^{RR}=5.9\pm0.8$
kpc$^{-3}$ once the measured number density is increased to account for
completeness.

Applying the above procedure to Oo I RRab stars we obtain $q=0.59\pm0.05$,
$n=2.4\pm0.09$, and $\rho_\sun^{RR}=4.0\pm0.7$ kpc$^{-3}$
($\rho_\sun^{RR}=5.0\pm0.9$ kpc$^{-3}$ when corrected for completeness). For the
Oo II subsample, we obtain $q=0.56\pm0.07$, $n=3.1\pm0.2$, and
$\rho_\sun^{RR}=2.1\pm1.0$ kpc$^{-3}$ ($\rho_\sun^{RR}=2.6\pm1.3$ kpc$^{-3}$
when corrected for completeness). The uncertainties on these parameters were
adopted from the analysis of mock Oo I and Oo II RRab samples. The ``height''
maps for the Oo I and Oo II RRab subsamples are shown in
Figure~\ref{Oo_peak_maps}. These power-law indices are consistent with indices
obtained by \citet{mic08} for LONEOS-I Oo I and Oo II RRab subsamples (see their
Table 3).

\subsection{Rejection of a Simple Power-Law Model}\label{variable_powerlaw}

The best-fit number density model of the full RRab sample ($n=2.4$, $q=0.63$) is
consistent with previous models for the number density profiles within 30 kpc
from the Galactic center of i) RR Lyrae stars ($n=2.4$ assuming $q=1.0$,
\citealt{wat09}; $n=2.3$ assuming $q=0.64$, \citealt{ses10a}), ii) metal-poor
main-sequence stars ($n = 2.6$ and $q = 0.7$, \citealt{sji11}), and
iii) blue horizontal branch stars ($n = 2.4$ and $q = 0.6$, \citealt{dea11}).
However, a closer inspection of $\Delta\log\rho$ residuals in the $R$ vs.~$Z$
map (top left panel in Figure~\ref{RZ_maps}), indicates that the best-fit single
power-law model overestimates the observed number density of RRab stars for
$r<16$ kpc. As shown in Figure~\ref{fit_check} (thick solid line), the model at
$r\sim5$ kpc predicts about 3 times more stars than observed.

A possible explanation for this discrepancy could be the incompleteness of the
RRab sample at distances closer than 16 kpc. However, this is unlikely as
Figure~\ref{completeness_plot} shows the completeness of RRab stars to be about
$\sim80\%$ within 20 kpc. Adding RR Lyrae stars from the LINEAR Catalog of
Variable Stars (see the end of Section~\ref{final_estimation}) does not
alleviate this discrepancy. Another possible explanation is an inadequate model
for number density variation with position. 

Two simple model modifications include variable oblateness parameter $q$, and a
variable power-law index. The former is not supported by data: when two
subsamples selected by $R_{gc}<12$ kpc and $R_{gc}>12$ kpc are fit separately,
the best-fit $q$ is essentially unchanged. This invariance of oblateness with
distance is consistent with several recent studies \citep{mic08, sji11, dea11}.

A double power-law model is one possible extension of the single power-law model
used above. When a double power-law model with oblateness parameter
$q=0.65\pm0.03$, inner power-law index $n_{inner} = 1.0\pm0.3$, outer power-law 
index of $n_{outer} = 2.7\pm0.3$, and a break radius at
$r_{br} = \sqrt{X^2+Y^2+(Z/q)^2}=16\pm1$ kpc is fit to data, the residuals
improve, but the model now underestimates the observed number densities within
$r\lesssim10$ (see Figures~\ref{RZ_maps} and~\ref{fit_check}). Again, the errors
on best-fit parameters are statistical uncertainties adopted from the analysis
of mock RRab samples, as done in the previous Section. The remaining residuals
within $r\lesssim10$ may be (again) due to an inadequate model or they may be
due to an overdensity (a diffuse overdensity has been reported in this region,
see Figure~\ref{XY_map_Virgo}). A power-law with an index of $n=2.5$ can model
the residuals within $r\lesssim10$ fairly well, but that would mean that the
number density distribution of RR Lyrae stars in the halo has a much more
complicated shape than ever reported (three power-laws).

Another possible explanation for the peculiar shape of the observed number
density distribution is that the halo does not have a smooth distribution
of RR Lyrae stars, with an occasional overdensity imprinted on top of it, but
that the distribution is more clumpy. To verify this hypothesis, we generated a
mock sample consisting of 400 uniform spheres with a radius of 2 kpc, with each
sphere containing 7 stars. The residuals obtained after fitting a single
power-law model to the number density distribution of this clumpy mock sample
are shown in Figure~\ref{RZ_maps} (bottom right panel) and
Figure~\ref{fit_check} (dotted line). As evident from Figure~\ref{fit_check},
the residuals for the clumpy mock sample and the observed sample exhibit similar
trends with $\log r$ when fitted with a single power-law. The agreement is
qualitative enough to suggest that the shape of the observed number density
distribution of RRab stars could be due to a purely clumpy halo.

We have used a broken power law to separately model the number densities of the
Oo I and Oo II samples. For the former, the best-fit parameters are the same as
for the full sample (recall that Oo I stars account for 75\% of the full
sample). The best parameters for the Oo II sample are $q=0.60$, $n_{inner}=1.6$,
$n_{outer} = 3.4$, $r_{br} = 18$ kpc, and $\rho_\sun^{RR}=0.6\pm0.5$ kpc$^{-3}$.
Therefore, for both single and broken power-law models, the number density
distribution for the Oo II subsample is steeper than for the Oo I subsample. 

\section{Halo Substructures}\label{substructure}

In the previous Section, we used the spatial distribution of LINEAR RRab stars
to estimate the best-fit smooth model for their number density distribution. In 
this Section, we use a group-finding algorithm {\em EnLink} \citep{sha09} to
identify significant clusters of RR Lyrae stars (halo substructures). The search
for halo substructures is done using LINEAR RRab stars that pass
Equations~\ref{spatial_cuts1}--\ref{spatial_cuts2}. The sample of RRab stars is
not split by Oosterhoff type so that groups that are
``Oosterhoff intermediate'', i.e., groups that may be associated with remnants
of dSph galaxies, can be detected as well. 

The algorithm EnLink has two free parameters which need to be supplied by the
user. The first parameter is the number of nearest neighbors employed for
density estimation, $k_{den}$. \citet{sha09} find that $k_{den} = 30$ is
appropriate for most clustering tasks, and we adopt their value. The second
parameter is the significance threshold $S_{th}$. Clusters of points that have
significance $S$ below threshold $S_{th}$ are denied the status of a group and
are merged with the background.

\citet{sha09} define the statistical significance $S$ for a group as a ratio of
signal associated with a group to the noise in the measurement of this signal.
The contrast, $\ln(\rho_{max})-\ln(\rho_{min})$ between the peak density of a
group ($\rho_{max}$) and valley ($\rho_{min}$) where it overlaps with another
group can be thought of as the signal, and the noise in this signal is given by
the variance $\sigma_{\ln\rho}$ associated with the density estimator
($\sigma_{\ln\rho}=0.22$ for $k_{den} = 30$). Combining the definitions of
signal and noise then leads to
$S = (\ln(\rho_{max})-\ln(\rho_{min}))/\sigma_{\ln\rho}$.

Selecting the value of $S_{th}$ is not trivial. For values of
$S_{th}$ that are too low, EnLink may detect spurious groups (i.e., groups
produced by Poisson noise). On the other hand, a threshold that is too high
might miss real halo substructures.

To find the optimal choice of $S_{th}$ for our sample of LINEAR RRab stars, we
run EnLink on ten mock samples of RRab stars drawn from a number density
distribution defined by Equation~\ref{RR_model}, where $n=2.42$, $q=0.63$, and
$\rho_\sun^{RR}=4.5$ kpc$^{-3}$ (the best-fit single power-law model; see
Section~\ref{number_density}). While this model may not be the best description
of the number density distribution of RRab stars in the halo (i.e., it
overestimates the number of RRab stars, see Section~\ref{variable_powerlaw}), it
is still useful. Mock samples drawn from this model will have a higher chance of
producing spurious groups (simply because they contain more stars), and thus the
number of spurious groups detected in such samples can serve as an upper limit
on the number of spurious groups that one may expect to find in the observed
sample.

To make mock samples similar to our sample of LINEAR RRab stars, we add noise to
heliocentric distances in mock samples and then trim down samples using
Equations~\ref{spatial_cuts1} to~\ref{spatial_cuts2}. The EnLink algorithm is
applied to each mock sample and the number of detected groups and the fraction
of stars in groups as a function of $S_{th}$ is recorded. The results are shown
in Figure~\ref{EnLink_threshold}.

When the significance threshold $S_{th}$ is low ($S_{th} < 0.7$), EnLink detects
several groups in mock samples. These groups are spurious, have low
significance, and they arise due to Poisson noise. As the threshold $S_{th}$
increases, the number of groups in mock samples decreases (i.e., random
fluctuations are not likely to create highly significant groups). The number of
groups detected in the observed sample also decreases with increasing $S_{th}$,
but at a different rate. For $S_{th} = 1.4$, EnLink detects seven groups in the
observed sample and on average one (spurious) group in mock samples. By
increasing the significance threshold to 2.5, we could eliminate the possibility
of detecting a spurious group in the observed sample, but the number of detected
groups would drop to two. We select $S_{th} = 1.4$ as the significance
threshold; this choice results in one expected spurious group.

The positions of detected groups, number of stars in a group, and significance
of a group are listed in Table~\ref{table2}. The spatial distribution of RRab
stars associated with these groups is shown in Figures~\ref{RRab_radec}
and~\ref{halo_sub_anim}. The stars associated with these groups are also
appropriately labeled in Table~\ref{table1} (see column ``Group ID''). On
average, the groups have radii of $\sim1$ kpc, where the radius is the median
distance of stars in a group from the peak in number density for that group.

\subsection{Detected Groups}\label{groups}

Of the seven groups, three groups (groups 3, 4, 5) are near globular clusters
(M53 or NGC 5053, M3, and M13), and one group (group 6) is located in the Virgo
constellation where several halo substructures have already been reported
\citep{viv01,new02,duf06,jur08}. The groups 3 and 4 are not isotropic and seem
to have a stream-like morphology (Figure~\ref{halo_sub_anim}). For example,
group 4, which is located near the globular cluster M3, extends $\sim4$
kpc roughly parallel to the Galactic plane, and makes $\sim45\arcdeg$ angle with
the Sun--Galactic Center line. Group 3, which is located near globular clusters
M53 and NGC 5053, on the other hand, extends more towards the Galactic plane.
Previous studies have detected tidal streams around NGC 5053 and M53
\citep{lpw06, chu10}, and no streams have been detected around M3 \citep{gj06}.

The seemingly non-isotropic distribution of RRab stars in group 5, which is near
globular cluster M13, is likely due to limited spatial coverage (the group is
close to the edge of the probed volume). However, even though this group is near
globular cluster M13, its relationship with the cluster is quite tenuous since
M13 is known to have a very small number of RR Lyrae stars \citep{cle01}.

Groups 1, 2, and 7 do not seem to be near any known halo substructures (e.g.,
the Sagittarius tidal streams), dSph galaxies or globular clusters. Globular
clusters M92 and M13 are the closest clusters to group 1, but they are still off
by $\sim3$ kpc in heliocentric distance. This offset is well outside the
uncertainty in distance, especially since the metallicities of clusters are
known and can be used to calculate the absolute magnitudes of RRab stars
potentially originating from clusters (using Equation~\ref{abs_mag}). Group 7
contains only ten RRab stars and has borderline significance ($S=1.42$). Thus,
this group is the one most likely to be spurious.

The ratio of Oosterhoff type II (Oo II) to Oosterhoff type I (Oo I) RRab stars
in a group may provide additional clues on the nature of detected groups. As
shown in Section~\ref{Oosterhoff}, this ratio is 1:4 for the halo. For group 4,
this ratio is close to zero (0.05; see Table~\ref{table2}), indicating that the
group 4 is dominated by Oo I RRab stars. This result is interesting because M3
is classified as an Oo I-type globular cluster \citep{cat09} and is located
within group 4 (see bottom panel in Figure~\ref{period_vs_amp_M3}). Thus, the
fact that the group 4 and the globular cluster M3 have the same Oosterhoff type
may indicate their common origin.

Group 3 is near globular clusters NGC 5053 and M53, both of which are classified
as Oo II-type globular clusters \citep{cat09}. Again, we find many more Oo II
RRab stars from this group near the centers of globular clusters NGC 5053 and
M53 (bottom panel in Figure~\ref{period_vs_amp_M53}), and  the group as a whole
has a higher ratio of Oo II to Oo I RRab stars (0.79). The reason we are not
detecting more Oo II RRab stars and the reason this ratio is not much higher for
this particular group is due to the incompleteness of the LINEAR RRab sample in
crowded regions (e.g., near centers of globular clusters).

Group 3 may actually consist of two groups of stars that were joined by EnLink
into a single group. Looking at group 3 in Figure~\ref{halo_sub_anim}, we can
discern a ``stream'' of points that is parallel to the Galactic plane, and spans
$\sim4$ kpc at $\sim13.5$ kpc above the Galactic plane. The ratio of Oosterhoff
II to I types in this subgroup is 1:2, or higher than in the halo. Based on the
morphology of this subgroup and its distance from the globular clusters M53 and
NGC 5053 ($\sim2-3$ kpc), this subgroup may be a separate halo substructure that
EnLink joined with the NGC 5053/M53 group of RRab stars.

It is worth mentioning that although M53 and NGC 5053 are relatively close to
each other in space, they have radial velocities that differ by more than 100 km
s$^{-1}$ (\citealt{har96} catalog, 2010 edition). Followup spectroscopic studies
should take advantage of this fact when associating group 3 and its parts with
either of these globular clusters.

In principle, the proper motions of clusters could be used to identify RR Lyrae
stars that were tidally stripped, as such stars would follow the motion of the
clusters. In the case of M53 and NGC 5053, there are a few RR Lyrae stars near
$R.A.\sim196\arcdeg$ and $Dec\sim19\arcdeg$ (see the bottom panel of
Figure~\ref{period_vs_amp_M53}) that roughly align with the proper motion vector
of M53 and with the tidal stream reported by \citet{lpw06}. These stars may have
been tidally stripped and may be trailing M53 or NGC 5053. However, since the
proper motion of M53 is quite uncertain
($\mu_\alpha\cos\delta=0.5\pm1.0$ mas yr$^{-1}$, $\mu_\delta =-0.1\pm1.0$ mas
yr$^{-1}$; \citealt{ode97}), and since the proper motion of NGC 5053 has not yet
been reported, it is difficult to judge whether this is truly the case. In the
case of M3, the RR Lyrae stars in group 4 spread in the east-west direction (see
the bottom panel of Figure~\ref{period_vs_amp_M3}), or almost perpendicular to
the proper motion vector of M3 ($\mu_\alpha\cos\delta=-0.06\pm0.3$ mas
yr$^{-1}$, $\mu_\delta =-0.26\pm0.3$ mas yr$^{-1}$; \citealt{wwc02}). While this
observation could be used as an argument against the claim that the extended
parts of group 4 share a common origin with the globular cluster M3, we do point
out that the measured proper motion of M3 is still quite uncertain.

Group 6, which is located in the Virgo constellation where several halo
substructures have been reported so far, has the ratio of Oosterhoff types that
is consistent to the one found for the halo (0.23 vs.~0.25).

Groups 1 and 2 have Oosterhoff types ratios that are higher by a factor of 2 and
3, respectively, relative to the halo. Assuming that the number density
distributions of Oo I and Oo II RRab stars in the halo are the same, the
probabilities of drawing groups 1 and 2 are 0.07 and 0.01. For comparison, the
probability of drawing group 6 (in the Virgo constellation) is 0.21, 0.28 for
group 7, and 0.23 for the subgroup of group 3.

\section{Discussion and Conclusions}\label{sec:disc} 

This paper is the second one in a series based on light curve data collected by
the asteroid LINEAR survey. In the first paper, \citet{ses11} described the
LINEAR survey and photometric recalibration based on SDSS stars acting as a
dense grid of standard stars. Here, we searched the LINEAR dataset for variable 
RR Lyrae stars and used them to study the Galactic halo structure and
substructures.

While this paper was in preparation, a study of RR Lyrae stars selected from the
Catalina Surveys Data Release 1 (CSDR1) was announced \citep{dra13}. The two
works are largely complementary in terms of science; while Drake et al.~analyzed
radial velocity and metallicity distributions of CSDR1 RR Lyrae stars with
spectroscopic measurements from SDSS and focused on the Sagittarius stream,
we searched for new halo substructures and studied the number density
distribution of RR Lyrae stars as a whole and by Oosterhoff type.

In total, we have selected 4067 type $ab$ and 834 type $c$ RR Lyrae stars from
the recalibrated LINEAR dataset. These stars probe $\sim$8000 deg$^2$ of sky to
30 kpc from the Sun. The LINEAR sample of RR Lyrae stars has low contamination
($\sim$2\%) and is $\sim$80\% complete to 23 kpc from the Sun. To facilitate
follow-up studies, the coordinates and light curve properties (amplitude,
period, etc.) of stars in this sample are made publicly available, as well as
their Oosterhoff classification and whether they are associated with a halo
substructure.

We also provide light curve templates that were derived from phased light curves
of $\sim$400 bright LINEAR RRab and RRc stars. Even though we used these
templates to obtain more accurate model light curves of LINEAR RR Lyrae stars,
we emphasize that we did not attempt to prune the template set by averaging
templates with similar shapes (as done by Ses10), and do not suggest that this
new template set should replace the light curve template set constructed by
Ses10. However, we do provide the new templates to support future work at
extending RR Lyrae template light curves (templates are provided as
supplementary data in the electronic edition of the journal).

\begin{figure}
\epsscale{0.6}
\plotone{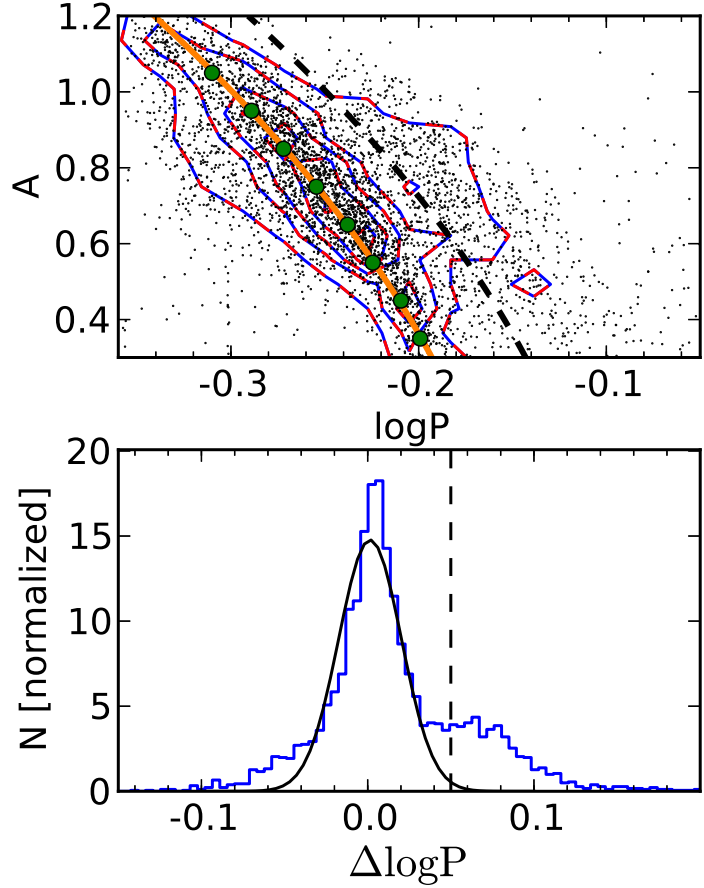}
\caption{
The points in the top panel show the distribution of LINEAR RRab stars in the
amplitude vs.~period diagram. The linearly spaced contours show 15\% to 75\% of
the peak density. The solid line shows the position of the Oo I locus, while the
dashed line (offset by $\Delta \log P = 0.05$ from the Oo I locus line)
separates the Oo I (to the left) and Oo II RRab stars (to the right). The Oo I
locus was obtained by fitting a quadratic line to the median of log periods
binned in narrow amplitude bins ({\em solid circles}). The position of the
dashed line was determined from the $\Delta \log P$ histogram ({\em bottom}),
where $\Delta \log P$ is the distance (at constant amplitude) from the Oo I
locus line. The Gaussian curve in the bottom panel models the peak associated
with Oo I RR abs stars, and the vertical dashed line in the bottom panel,
centered at $\Delta \log P = 0.05$, tentatively separates Oo I and Oo II RRab
stars in this histogram. The short-period tail of the $\Delta \log P$ histogram
likely contains Bla\v{z}ko RRab stars. The amplitudes of Bla\v{z}ko RRab stars
can be underestimated if they are not observed near the peak of their Bla\v{z}ko
cycle (when the light curve amplitude is highest), causing them to scatter
downwards in this diagram (i.e., towards lower amplitudes at constant periods).
\label{period_vs_amp}}
\end{figure}

\begin{figure}
\epsscale{1.0}
\plotone{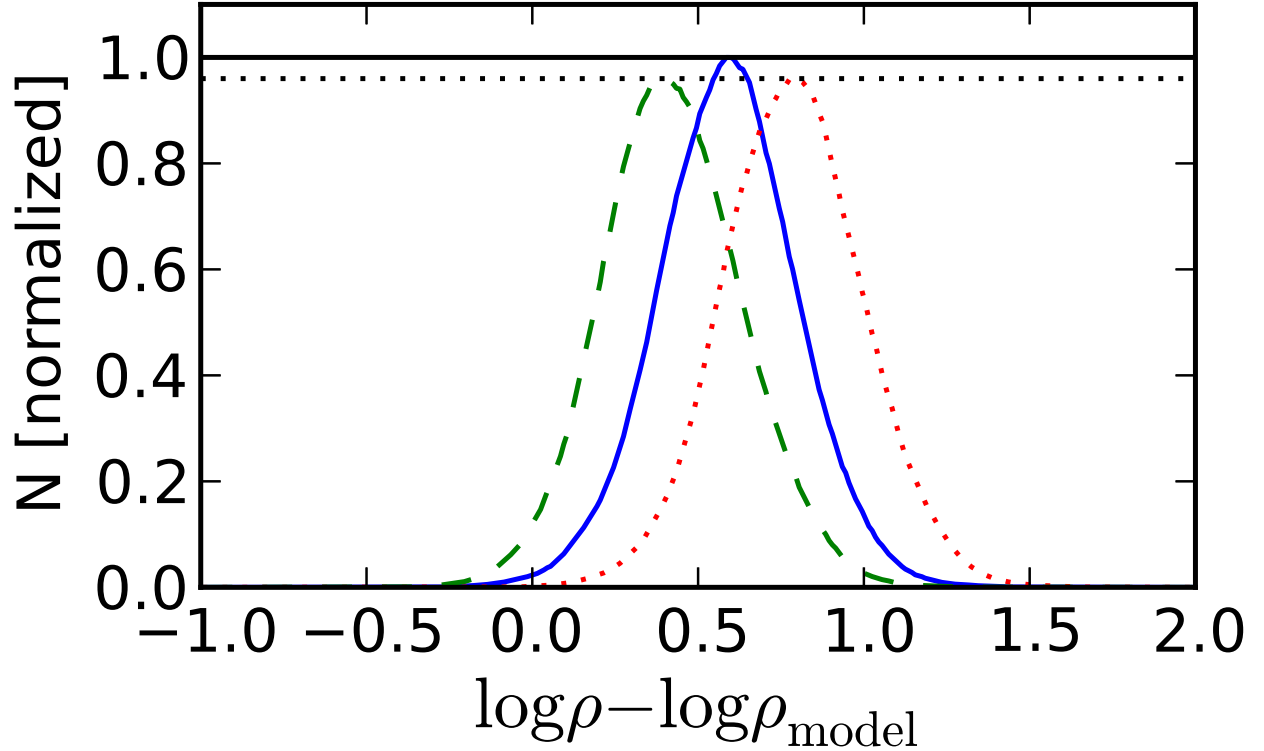}
\caption{
A comparison of $\Delta\log\rho$ histograms obtained with the correct number
density model ($q=0.71$, $n=2.62$; solid), a model with the wrong power-law
index $n$ ($q=0.71$, $n=3.1$; dotted), and a model with the wrong oblateness
parameter $q$ ($q=0.9$, $n=2.62$; dashed). The histograms are normalized to the
height of the histogram for the correct model. Note that the height of
histograms obtained with incorrect models (dashed and dotted) are lower than the
height of the histogram obtained with the correct number density model (solid). 
The difference in heights, illustrated by the horizontal solid and dotted line,
is 4\%. For a given $q$ and $n$, $\rho_\sun^{RR}$ is estimated as the mode of a
$\Delta\log\rho$ histogram raised to the power of 10. For example, for the
correct model $\rho_\sun^{RR} = 10^{0.65}\sim4.5$ kpc$^{-3}$.
\label{logdRho_hists}}
\end{figure}

\begin{figure}
\epsscale{1.0}
\plottwo{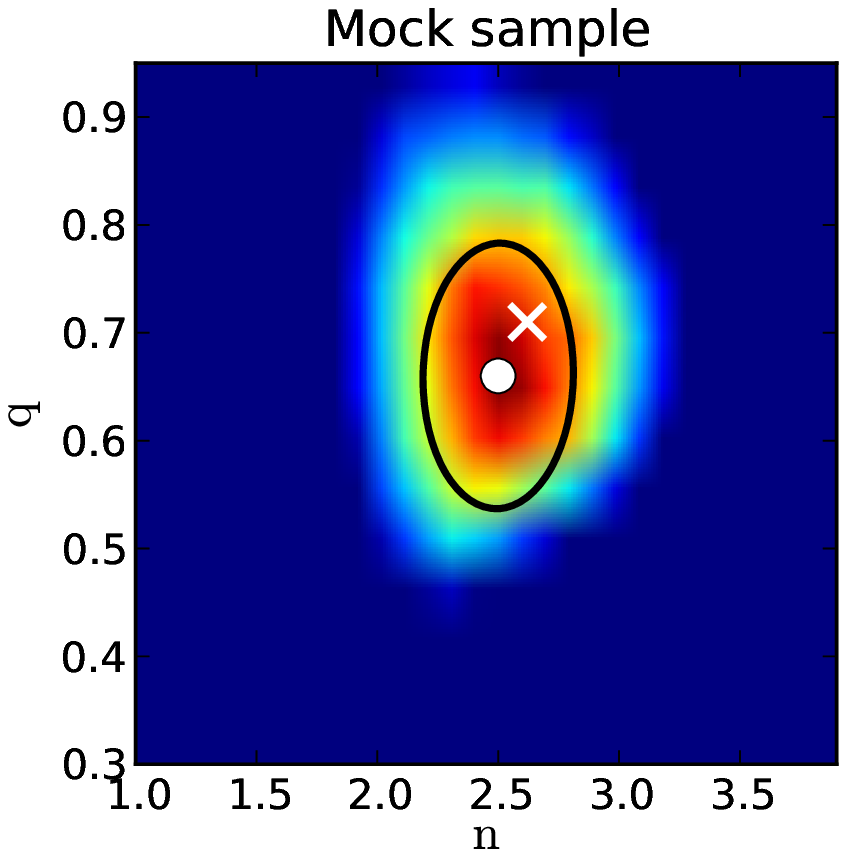}{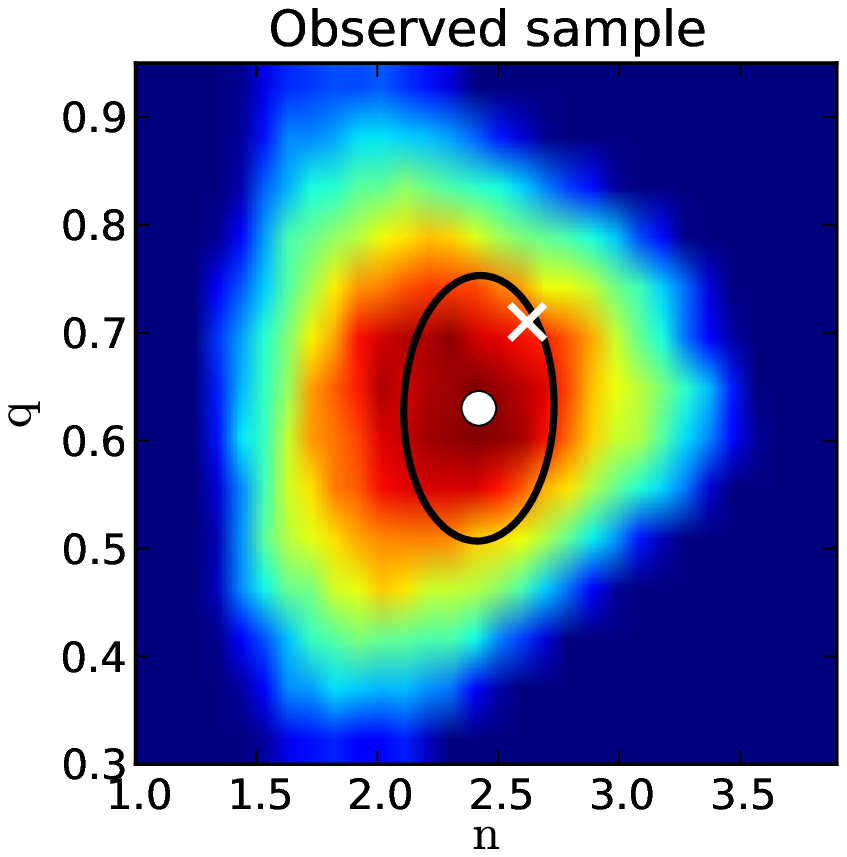}
\caption{
{\em Left}: A map showing dependence of the height of the $\Delta\log\rho$
histogram on the assumed oblateness parameter $q$ and power-law index $n$, for
one of the mock RR Lyrae samples generated using Galfast. The color indicates
the height of the $\Delta\log\rho$ histogram, with the red color representing
the greatest height. The position of the cross symbol indicates the $q$ and $n$
values used by \citet{sji11} to describe the halo stellar number density within 
30 kpc from the Sun. These values ($q=0.71$, $n=2.62$) were also used to create 
mock samples with Galfast. The position of the solid circle indicates the
best-fit obtained for this particular mock sample ($q=0.66$, $n=2.5$). At this
position, the height of the $\Delta\log\rho$ histogram is the greatest. The
best-fit and input values are consistent within the 95\% confidence limit
(ellipse). {\em Right}: A height map obtained using the sample of RRab stars
observed in LINEAR. The best-fit values of $q=0.63$ and $n=2.42$ are consistent
with \citet{sji11} values within the 95\% confidence limit (ellipse).
\label{peak_maps}}
\end{figure}

\begin{figure}
\epsscale{1.0}
\plottwo{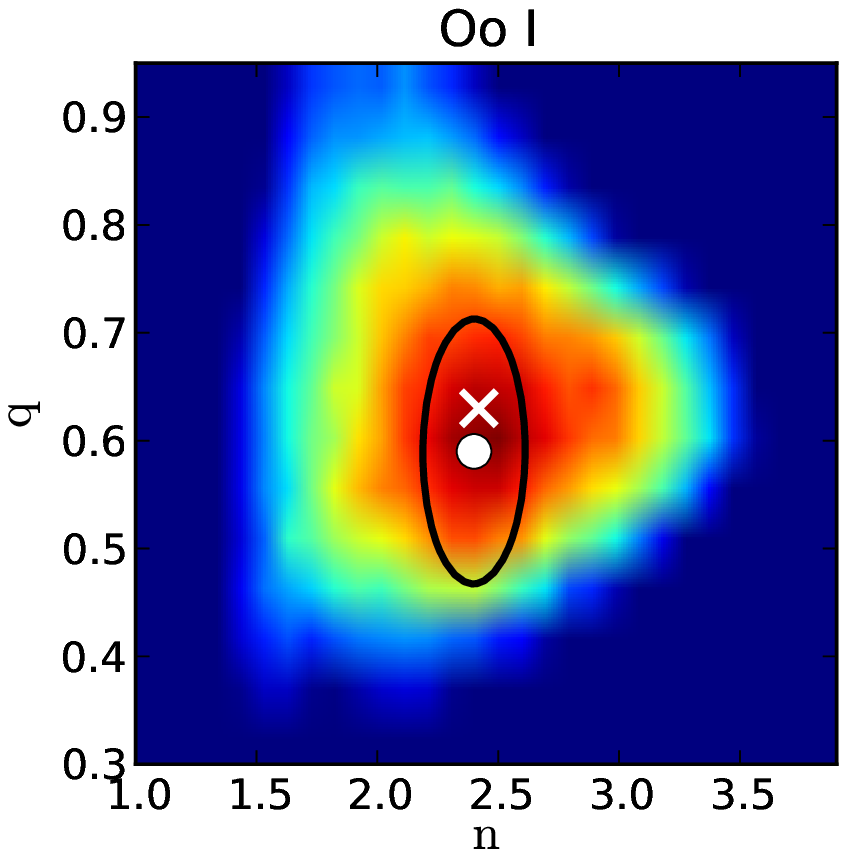}{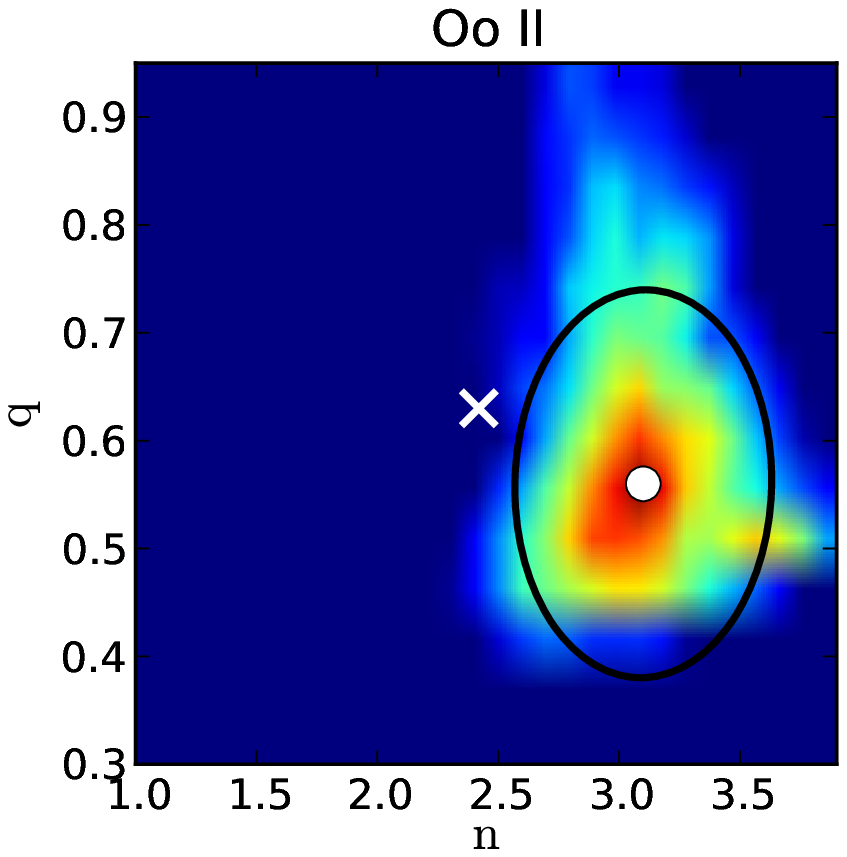}
\caption{
Color-coded ``height'' maps for the Oosterhoff type I ({\em left}) and type II
RRab subsamples ({\em right}). The ``x'' marks the best-fit parameters for the
full RRab sample ($q=0.63$, $n=2.42$), and the solid circle shows the best-fit
parameters for the Oo I ($q=0.59$, $n=2.4$) and Oo II subsamples ($q=0.56$,
$n=3.1$). The ellipse indicates the 95\% confidence limit.
\label{Oo_peak_maps}}
\end{figure}

\begin{figure}
\epsscale{0.8}
\plottwo{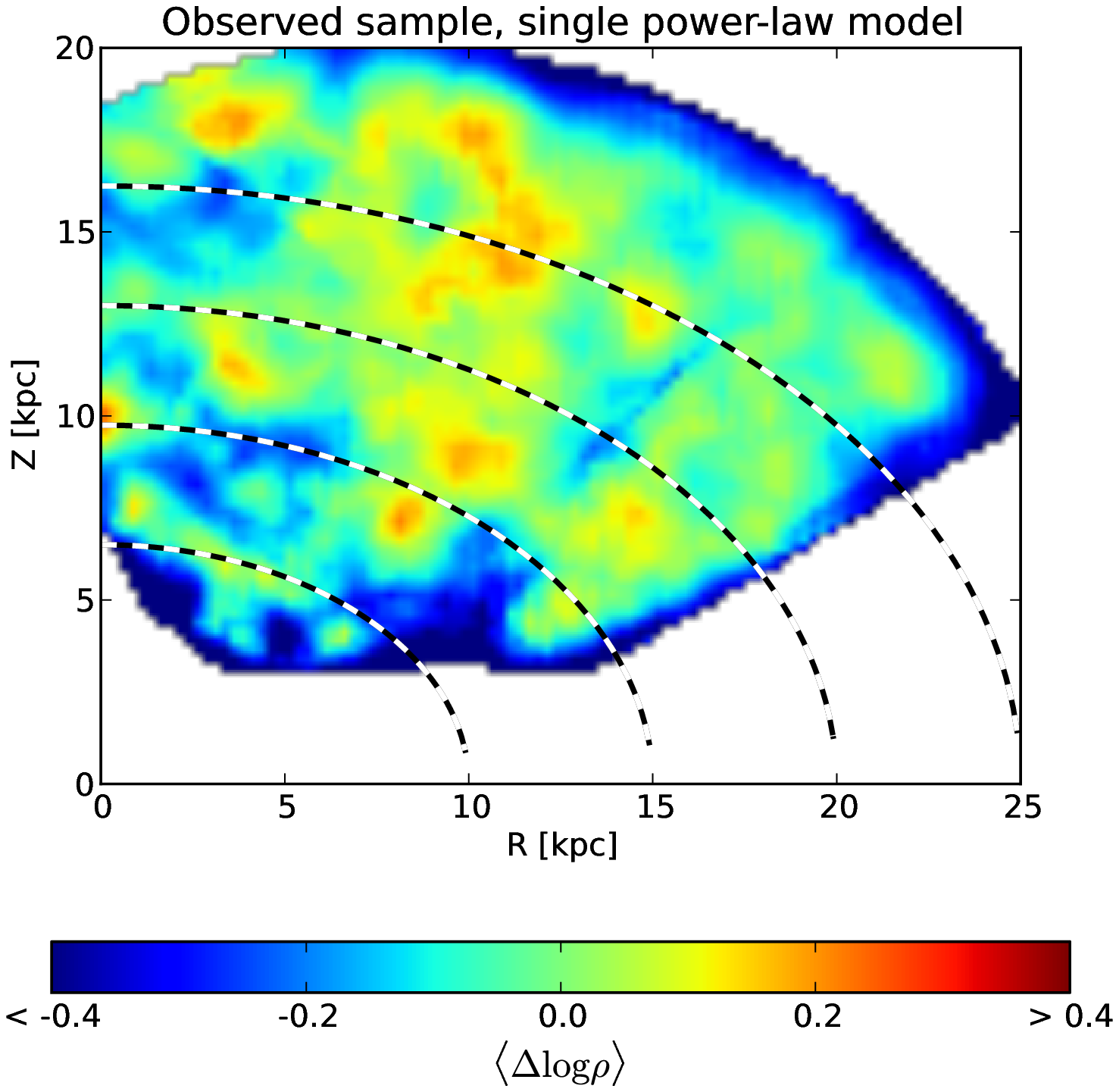}{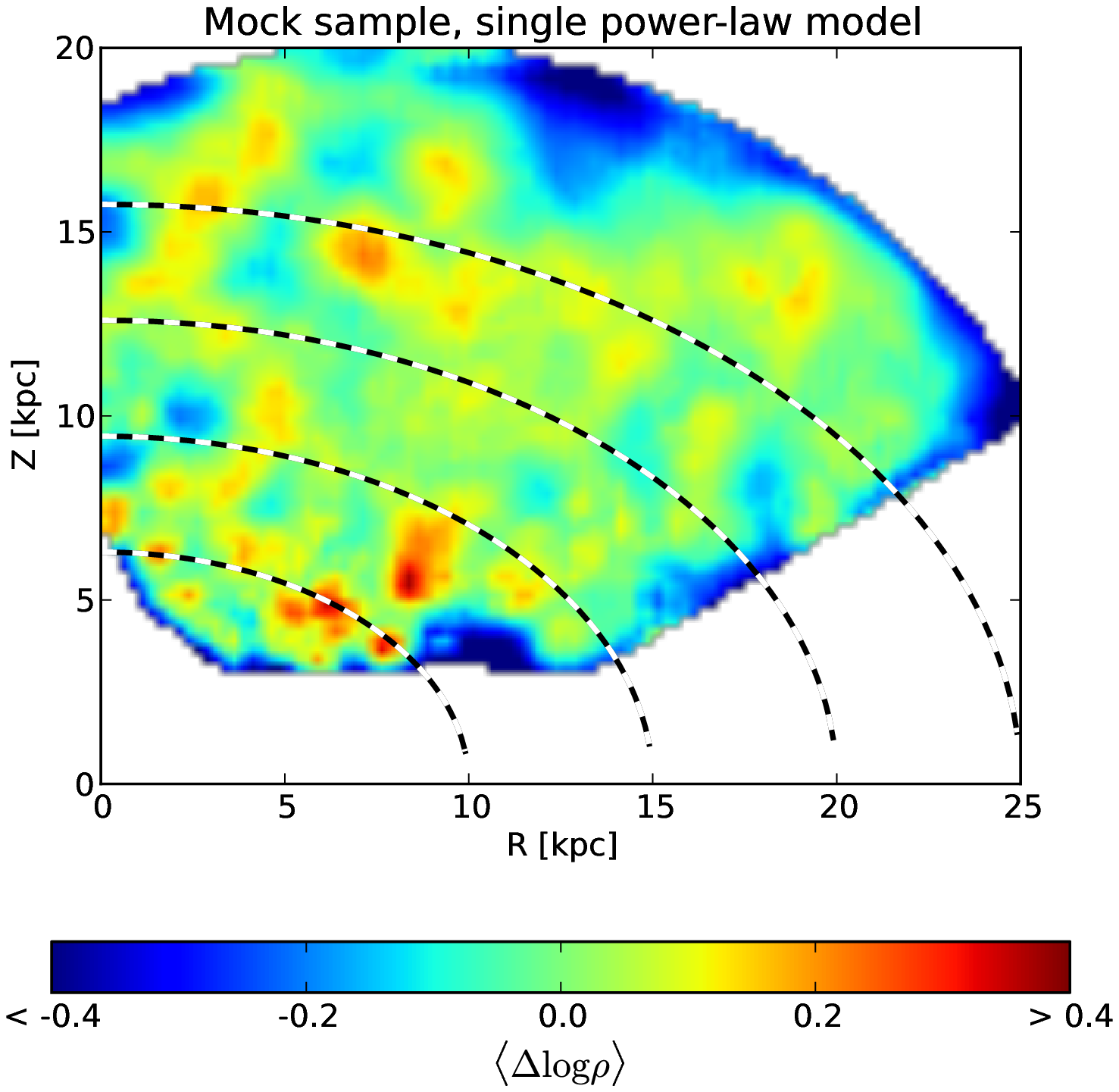}

\plottwo{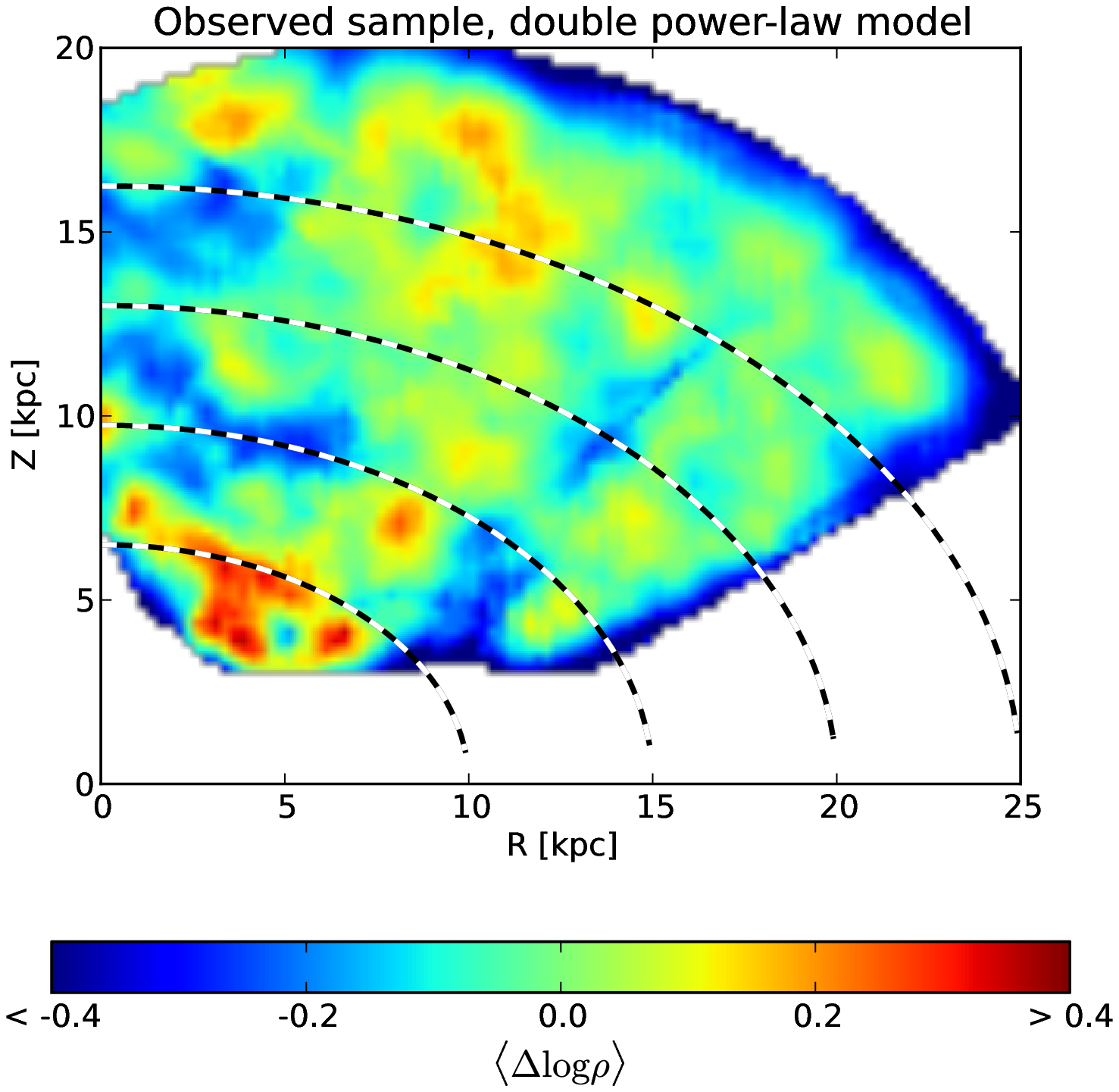}{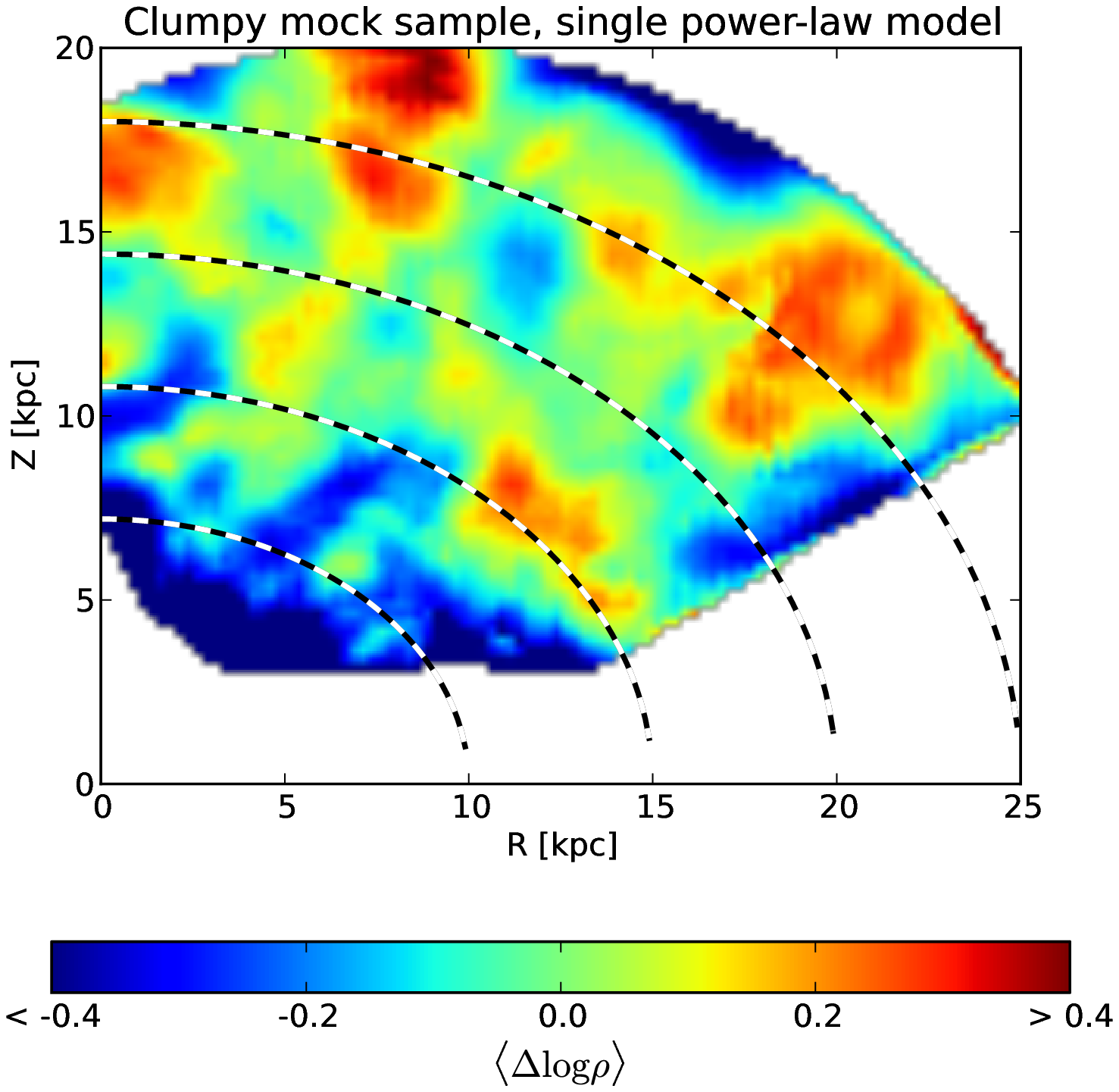}
\caption{
The top left $R=\sqrt{X^2 + Y^2}$ vs.~$Z$ map shows median $\Delta\log\rho$
residuals obtained by fitting a $q=0.63$, $n=2.42$ single power-law model to the
observed number density distribution of the full RRab sample. The residuals are
color-coded according to legend. The dashed lines show constant distances
$r=\sqrt{X^2 + Y^2 + (Z/q)^2}$ of 10, 15, 20, and 25 kpc. Note how this best-fit
single power-law model overestimates the number densities (median
$\Delta\log\rho < 0$) for $r<16$ kpc. For comparison, the top right panel shows
residuals obtained by fitting a single power-law to the number density
distribution of a mock sample drawn from a smooth $q=0.63$, $n=2.42$ model. The
residuals in this map illustrate the level of shot noise that is also present in
the observed sample (top left map). The bottom left map shows the residuals
obtained by fitting a double power-law to the observed number density
distribution of the full RRab sample. The residuals for $10<r/{\rm kpc}<16$ have
decreased, but the model now underestimates the number density within $r\sim10$
kpc. The bottom right map shows the residuals obtained by fitting a single
power-law to the number density distribution of a mock sample consisting of 400
uniform clumps (see Section~\ref{variable_powerlaw} for details). Note how the
residuals in this map are systematically more negative for $r\lesssim15$ kpc, a
trend that is also present in the observed sample (top left map).
\label{RZ_maps}}
\end{figure}

\begin{figure}
\epsscale{0.5}
\plotone{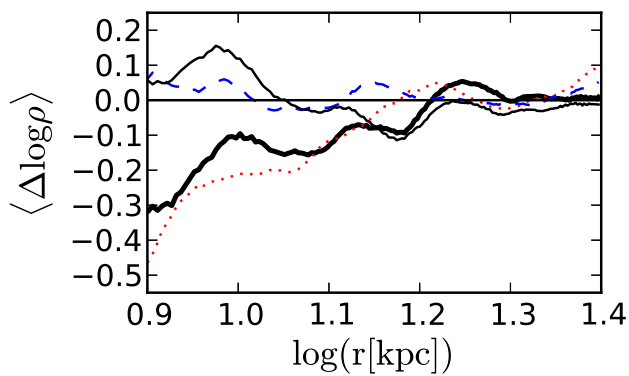}
\caption{
Dependence of median $\Delta\log\rho$ residuals shown in Figure~\ref{RZ_maps} on
log of distance $r=\sqrt{X^2 + Y^2 + (Z/q)^2}$. The thick solid line shows the
residuals for the top left map (observed sample fitted by a single power-law),
the dashed line shows the residuals for the top right map (smooth mock sample
fitted by a single power-law), the thin solid line shows the residuals for the
bottom left map (observed sample fitted by a double power-law), and the dotted
line shows the residuals for the bottom right map (clumpy mock sample fitted by
a single power-law).
\label{fit_check}}
\end{figure}

We find evidence for the Oosterhoff dichotomy among field RR Lyrae stars. While
we do not see the clearly displaced secondary peak in the $\Delta\log P$ diagram
(bottom panel in Figure~\ref{period_vs_amp}) that is usually associated with an
Oo II component (e.g., see Figure 21 by \citealt{mic08}), we do detect a long
tail that contains RRab stars with periods consistent with Oosterhoff type II
RRab stars. The ratio of the number of stars in the Oosterhoff II and I
subsamples is 1:4. A similar ratio was found by \citet{mic08} and \citet{dra13}
(0.26 and 0.24, respectively).

The lack of a clear gap in the $\Delta\log P$ distribution between the two
components may be a combination of two factors. First, this region in the
$\Delta\log P$ diagram may contain ``Oosterhoff intermediate'' RR Lyrae stars.
As these stars are presumed to come from dwarf galaxies similar to present-day
Milky Way dSph satellite galaxies \citep{cat09}, the lack of a gap may be
evidence that some fraction of stars were accreted from such systems. Second,
the gap may be filled by short-period Oosterhoff type II RR Lyrae stars that are
undergoing Bla\v{z}ko variations. If they are not observed at the maximum of
their Bla\v{z}ko cycle when their light curve amplitude is the greatest, or if
their maximum light curve amplitude is not recognized in the folded data, these
stars will scatter towards lower amplitudes at the same period and will fill in 
the gap between the two Oosterhoff components.

The wide coverage and depth of the LINEAR RRab sample allowed us to study the
number density distribution of halo RRab stars to a much greater extent than it
was possible in previous studies. We find it possible to describe the number
density of RRab stars by an oblate $1/r^{n}$ ellipsoid, with the axis ratio
$q = 0.63$ and the power-law index of $n=2.42$. These values are consistent with
previous models for the number density of various tracers within 30 kpc from the
Galactic center \citep{wat09, ses10a, sji11, dea11}.

\begin{figure}
\epsscale{1.0}
\plotone{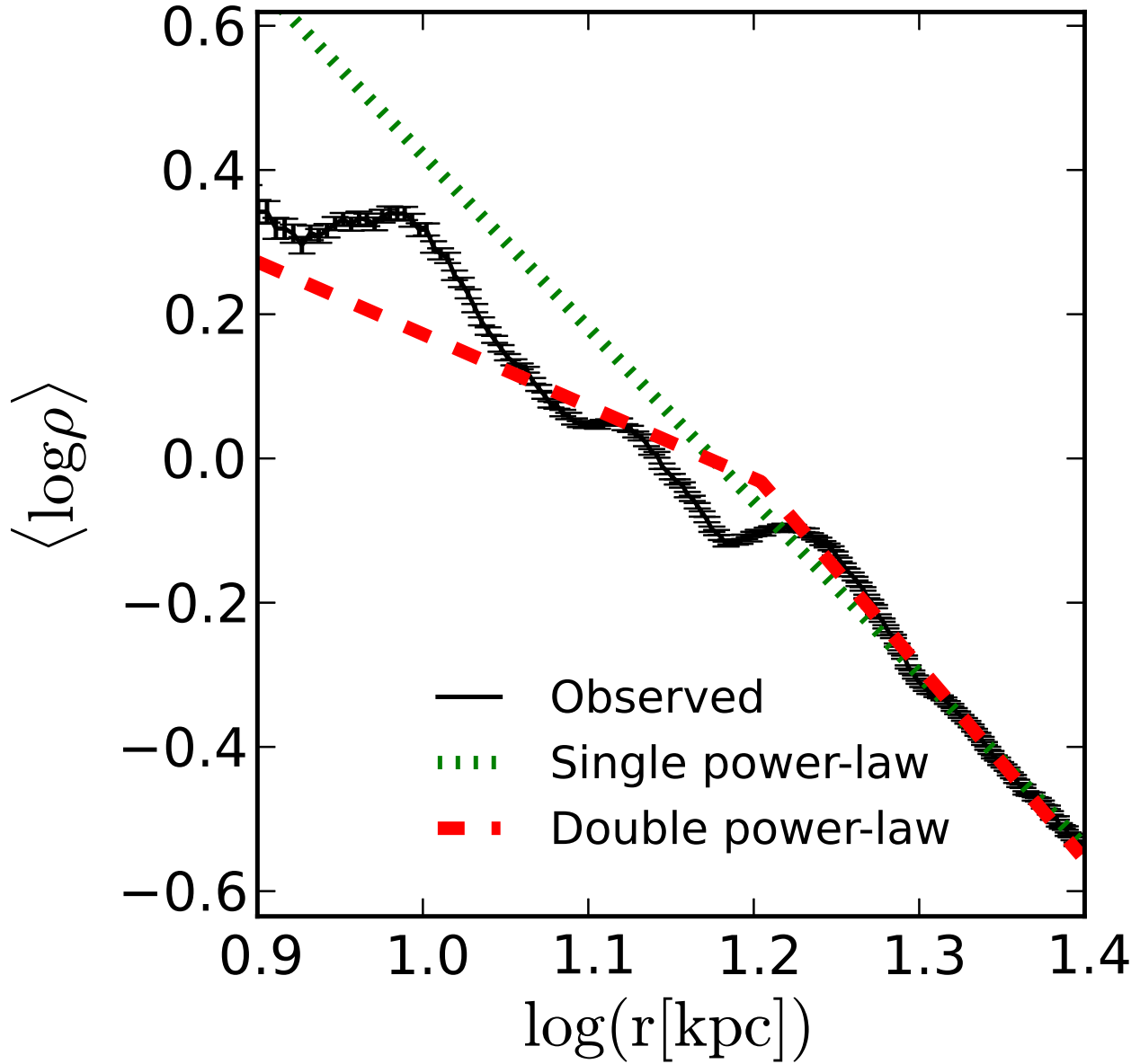}
\caption{
A comparison of the median observed and model log number densities as a function
of log distance $r$. The error bars show the error in the median observed
$\log\rho$. The best-fit single power-law model has oblateness parameter
$q=0.63$ and power-law index $n=2.42$, while the double (or broken) power-law
has oblateness parameter $q=0.65$ and a power-law index changing from
$n_{inner}= 1.0$ to $n_{outer} = 2.7$ at $r_{br} \sim 16$ kpc.
\label{logRho_logr_comparison}}
\end{figure}

\begin{figure}
\epsscale{1.0}
\plotone{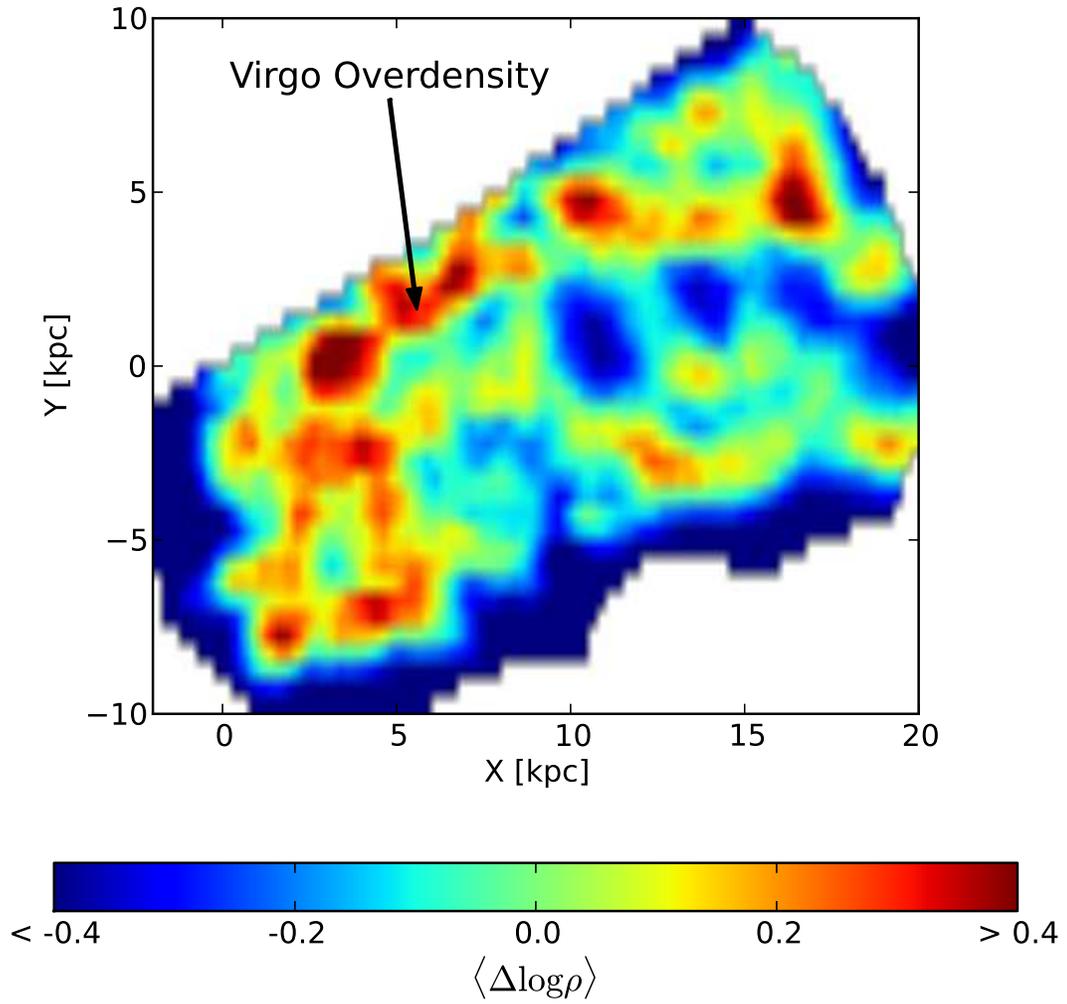}
\caption{
An $X$ vs.~$Y$ map showing median $\Delta\log\rho$ residuals in the $3 < Z < 7$
kpc range (the vicinity of the Virgo Overdensity; \citealt{jur08}). The
residuals come from a comparison with the best-fit double power-law model.
\label{XY_map_Virgo}}
\end{figure}

\begin{figure}
\epsscale{0.74}
\plotone{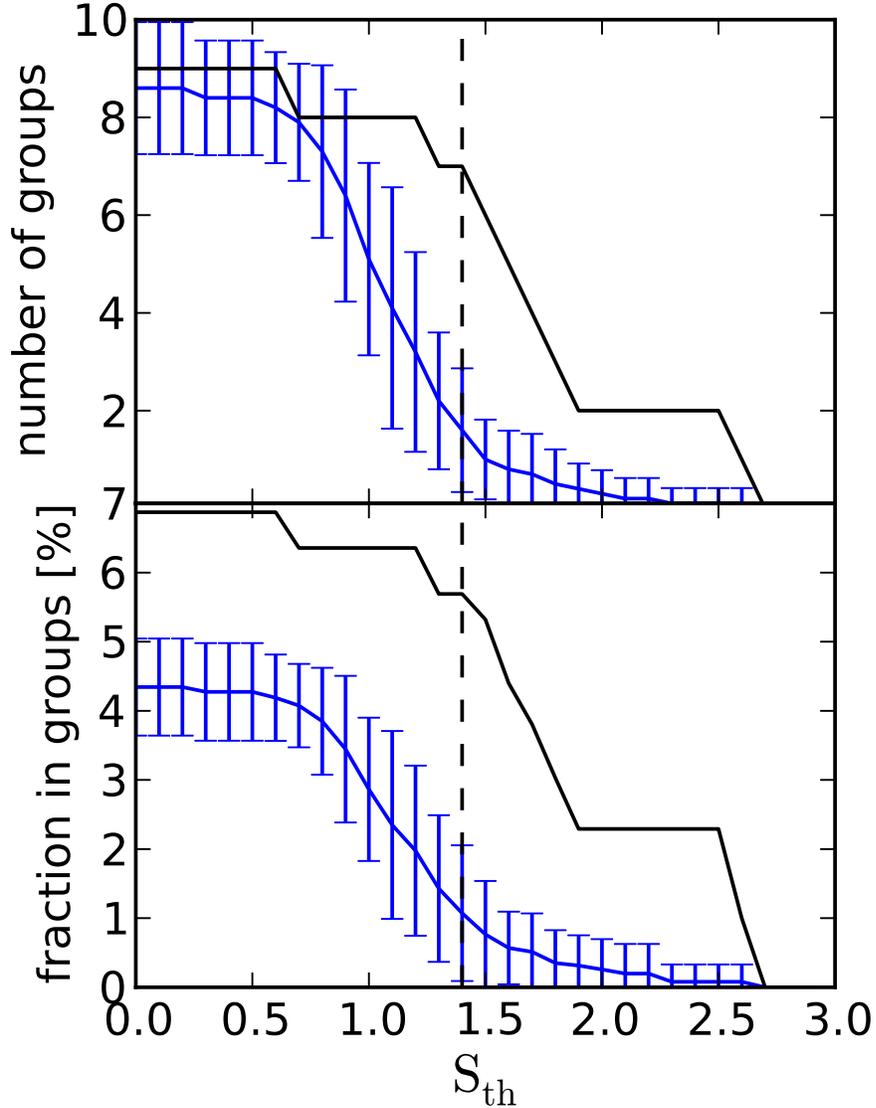}
\caption{
The number of groups ({\em top}) and the percentage of stars in groups
({\em bottom}) identified by EnLink as a function of significance threshold
$S_{th}$. The line with error bars shows the dependence obtained by running
EnLink on 10 mock samples of RR Lyrae stars that have no real groups. The error
bars show the standard deviation. The line without errorbars shows the
dependence obtained by running EnLink on the observed sample of LINEAR RRab
stars. For $S_{th}=1.4$ (dashed line), EnLink identifies seven groups in the
observed sample and one spurious group in mock samples. In the observed sample,
these seven groups contain about 5.5\% of all stars in the sample. In the mock
samples, the one spurious group contains about 1\% of all stars.
\label{EnLink_threshold}}
\end{figure}

\begin{figure}
\epsscale{1.0}
\plotone{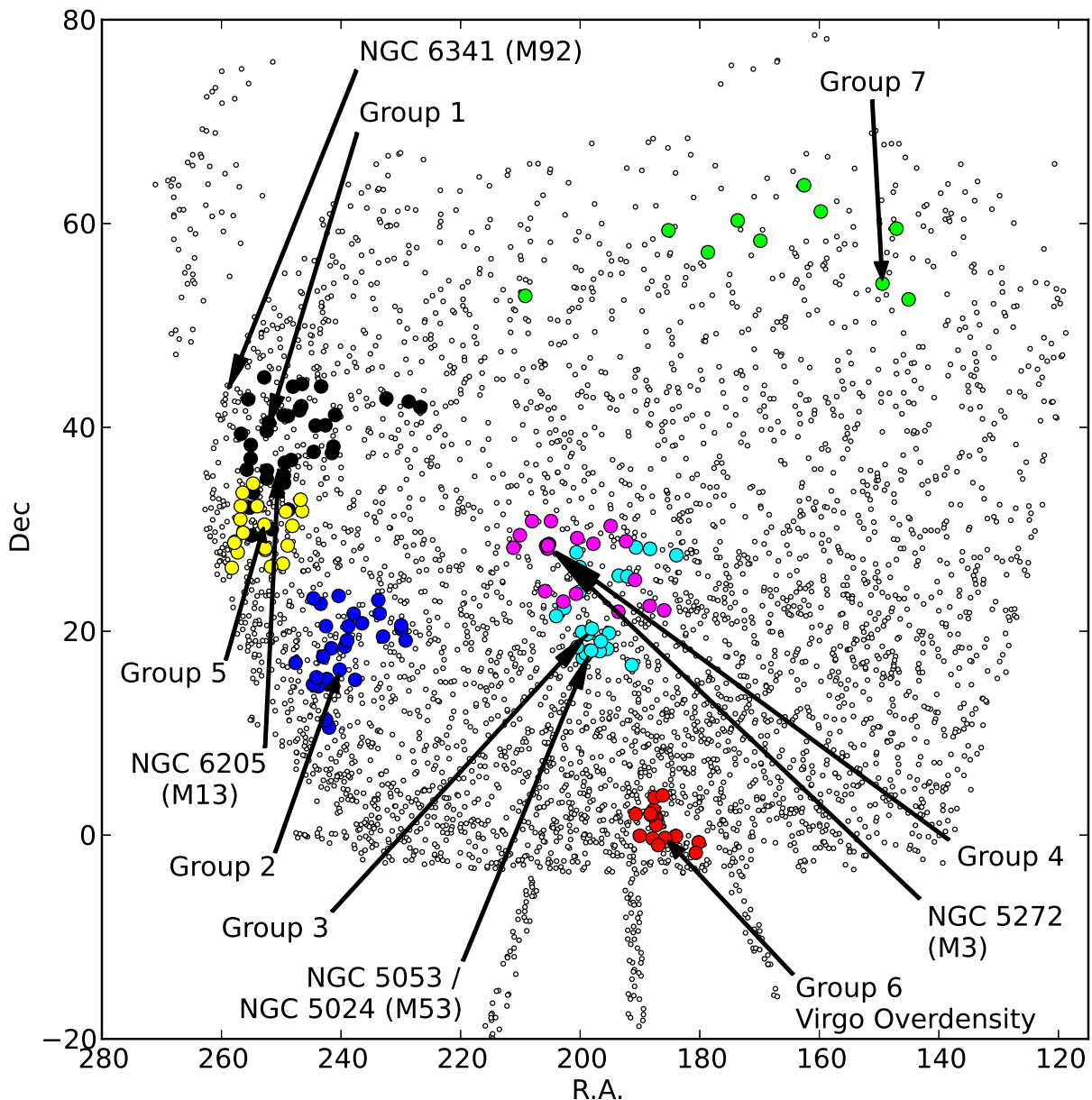}
\caption{
The spatial distribution of 4067 RRab stars selected from the LINEAR survey
(small open circles). The solid circles show positions of RRab stars associated
with significant groups and arrows point to positions of peaks in number
density. Positions of globular clusters in the vicinity of groups are indicated 
by arrows.
\label{RRab_radec}}
\end{figure}

\begin{figure}
\epsscale{1.0}
\plotone{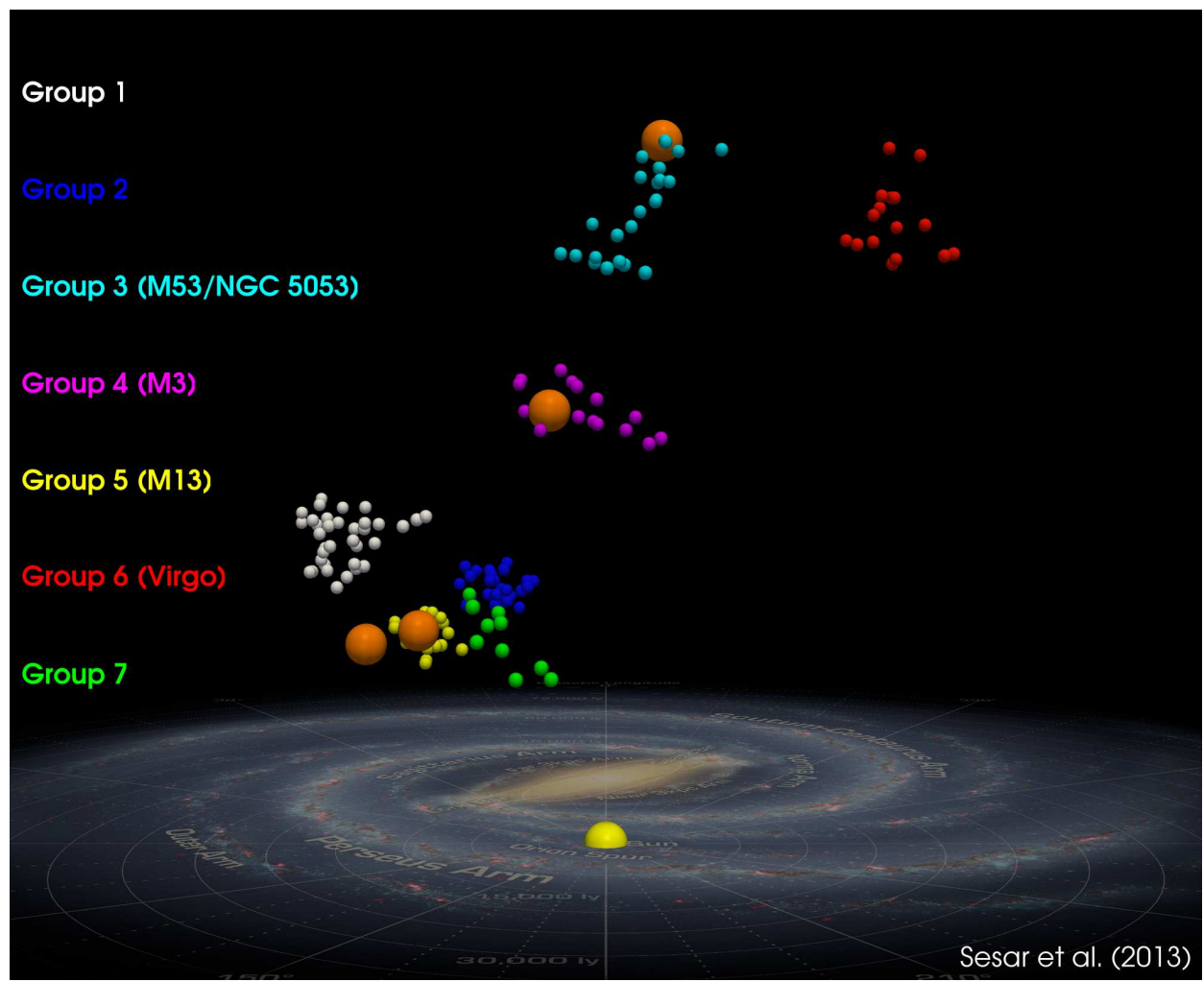}
\caption{
A single frame of an animation showing a fly-by of halo substructures detected
in this work and of the scaled Galactic plane (annotated artist's concept by
NASA/JPL-Caltech). The small spheres show LINEAR RRab stars associated with
significant groups and the larger spheres show the positions of globular
clusters in the vicinity of groups. The heliocentric distances of globular
clusters, which were taken from the \citet{har96} catalog (2010 edition), have
been multiplied by $10^{0.2(0.23{\rm [FeH]}+0.93 - 0.6)}$ to account for the
difference between the absolute magnitude of RRab stars in the cluster and the
one adopted for all RRab stars in this work ($M_{RR} = 0.6$; see
Section~\ref{distances}). Going from left to right, the globular clusters are
M92, M13, M3 and M53. No known halo substructures, globular clusters or dSph
galaxies are near groups 1, 2, and 7. Note the extended morphology of groups 3
and 4. These groups may trace tidal streams related to, or in the vicinity of,
M53 and M3 globular clusters. The animation is provided in the electronic
edition of the Journal.
\label{halo_sub_anim}}
\end{figure}

\begin{figure}
\epsscale{0.65}
\plotone{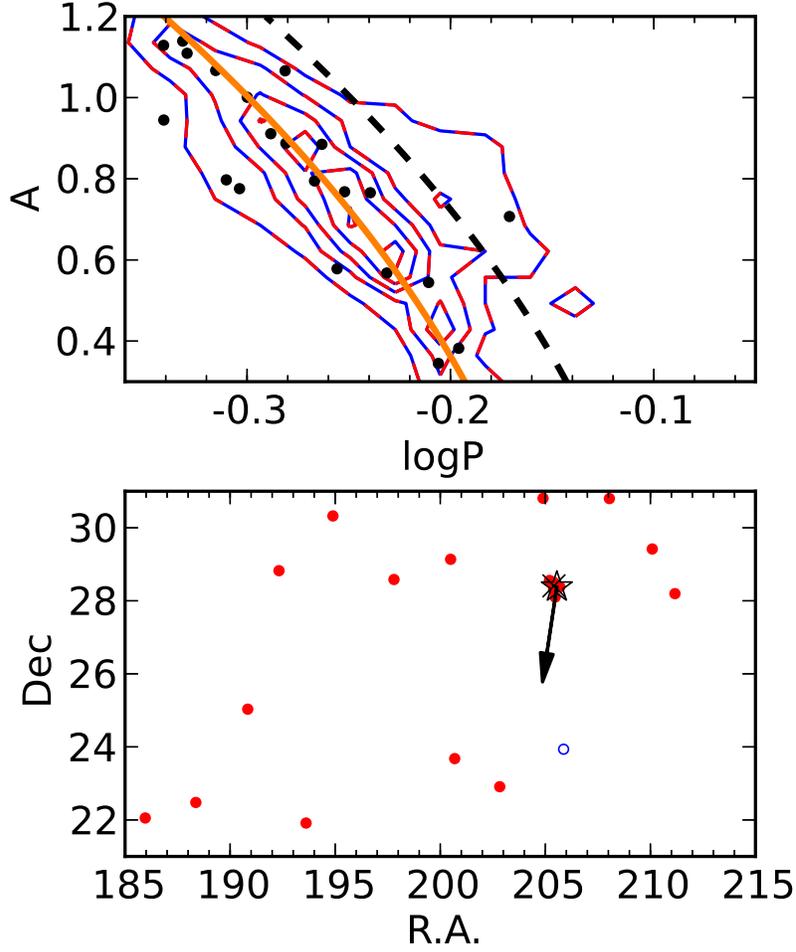}
\caption{
The distribution of RRab stars from group 4 in the amplitude vs.~period diagram
({\em top}) and right ascension vs.~declination plot ({\em bottom}). The
contours, solid, and dashed lines are the same as in Figure~\ref{period_vs_amp}.
The ``x'' symbol shows the position of the peak in number density and the
(overlapping) open star symbol shows the position of globular cluster M3
according to the \citet{har96} catalog (2010 edition). The arrow indicates the
proper motion of M3 ($\mu_\alpha\cos\delta=-0.06\pm0.3$ mas yr$^{-1}$,
$\mu_\delta =-0.26\pm0.3$ mas yr$^{-1}$; \citealt{wwc02}). Oosterhoff type I (Oo
I) and Oosterhoff type II (Oo II) RRab stars are shown as solid and open
circles, respectively. The 6 RRab stars located near the center of M3 are known
cluster members \citep{cle01}. In the bottom panel, the extended distribution of
RRab stars well outside the cluster's $30\arcmin$ tidal radius suggests presence
of a possible tidal stream.
\label{period_vs_amp_M3}}
\end{figure}

\begin{figure}
\epsscale{0.6}
\plotone{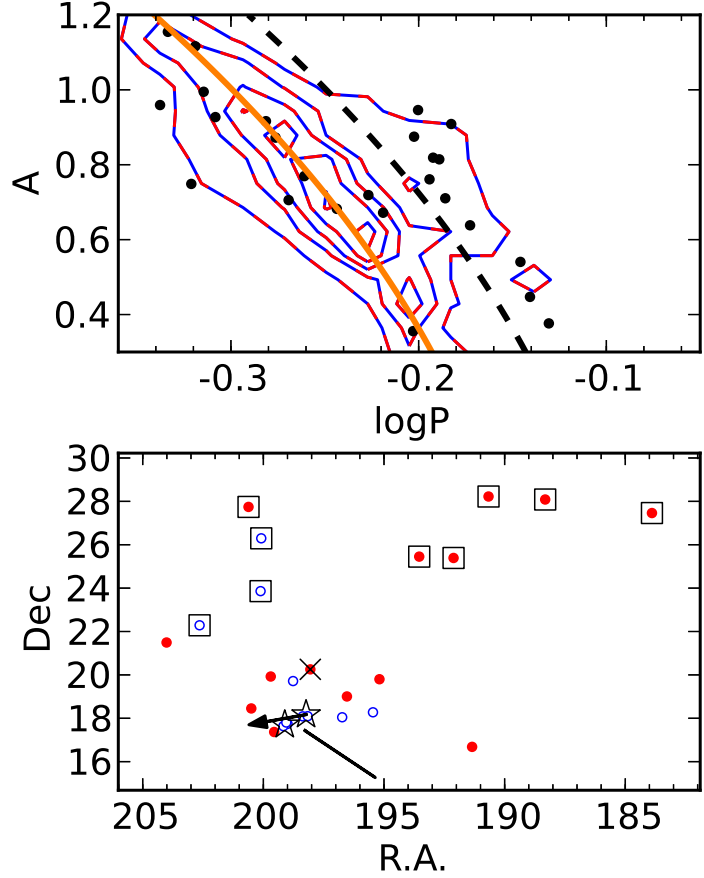}
\caption{
Similar to Figure~\ref{period_vs_amp_M3}, but for RRab stars from group 3. The
open star symbols shows the positions of globular clusters NGC 5053 (left) and
M53 (right) according to the \citet{har96} catalog (2010 edition). The solid
line shows the extent of the NGC 5053 tidal stream reported by \citet{lpw06},
and the arrow indicates the proper motion of M53
($\mu_\alpha\cos\delta=0.5\pm1.0$ mas yr$^{-1}$, $\mu_\delta =-0.1\pm1.0$ mas
yr$^{-1}$; \citealt{ode97}). The proper motion of NGC 5053 has not yet been
reported. Presence of a significant number of Oo II RRab stars in this group
(open circles) and their proximity to centers of Oo II-type globular clusters
M53 and NGC 5053 further strengthen the association of at least a part of group 
3 with these globular clusters. The circles enclosed in squares are stars
located between 13 and 14 kpc from the Galactic plane. These stars are evident
as a ``stream'' of points in group 3 that is parallel to the Galactic plane and
that spans $\sim4$ kpc (see Figure~\ref{halo_sub_anim}). The ratio of Oosterhoff
II to I types in this subgroup is 1:2. Based on its morphology, this subgroup
may be a separate halo substructure that EnLink joined with the NGC 5053/M53
group of RRab stars.
\label{period_vs_amp_M53}}
\end{figure}

However, as discussed in Section~\ref{variable_powerlaw} and as illustrated in
Figure~\ref{logRho_logr_comparison}, the single power-law model overestimates
the number density of RRab stars within $r\sim16$ kpc. Analysis done in
Section~\ref{completeness} excludes incompleteness as a likely explanation for
this discrepancy. Potential problems with the fitting method are also unlikely, 
as we have repeatedly tested our fitting method on several mock samples drawn
from known number density distributions, and have quantified its precision in
recovering input parameters in the presence of realistic distance errors and
halo substructures. The discrepancy between the best-fit single power-law model 
and the observed number density can be decreased by using a broken (or double)
power-law model, where the power-law index changes from $n_{inner}=1.0$ to
$n_{outer}=2.7$ at the break radius of $r_{br}\sim16$ kpc, with the best-fit
oblateness parameter set at $q=0.65$. The variation in the power-law index is
not due to the smaller Oosterhoff II component, because this subsample also
shows independent evidence for a variable power-law. Alternatively, as
simulations with mock samples have suggested, the observed variable power-law
slope may not due to a change in the smooth distribution of stars, but may be
evidence of a clumpy distribution of RR Lyrae stars within $r\sim16$ kpc.

Possible independent evidence that may support the density profile observed in
Figure~\ref{logRho_logr_comparison} may be found in a recent kinematic study by
\citet{kaf12}. Kafle et al.~used 4667 blue horizontal branch (BHB) stars
selected from the SDSS/SEGUE survey to determine key dynamical properties of the
Galactic halo, such as the profile of velocity anisotropy $\beta$
\begin{equation}
\beta = 1 - \frac{\sigma^2_\theta + \sigma^2_\phi}{2\sigma^2_r},
\end{equation}
where $\sigma_r$, $\sigma_\theta$, and $\sigma_\phi$ are the velocity
dispersions in spherical coordinates.
They find that ``from a starting value of $\beta\approx0.5$ in the inner parts 
($9 <r/{\rm kpc} <12$), the profile falls sharply in the range $r\approx13-18$
kpc, with a minimum value of $\beta=-1.2$ at $r = 17$ kpc, rising sharply at
larger radius''. The metal-rich and metal-poor population of BHB stars analyzed
by \citet{kaf13} were also found to exhibit similar behaviour in $\beta$. The
range of distances where the $\beta$ sharply falls is the same range where we
find that a shallower power-law with an index of 1.0 provides as better fit to
observed number densities of RRab stars. This suggests that the two effects may
be related to each other and may have a common cause (e.g., presence of a
diffuse substructure).

Using a group-finding algorithm {\em EnLink}, we searched for halo substructures
in our sample of RRab stars and detected seven candidate halo groups, one of
which may be spurious (based on a comparison with mock samples). Three of these
groups are near globular clusters (groups 3, 4, and 5), and one (group 6) is
near a known halo substructure (Virgo Stellar Stream; \citealt{viv01, duf06}).
The extended morphology and the position (outside the tidal radius) of some of the groups near globular clusters is suggestive of tidal streams possibly
originating from globular clusters. The remaining three groups do not seem to be
near any known halo substructures or globular clusters. Out of these, groups 1
and 2 have a higher ratio of Oosterhoff type II to Oosterhoff type I RRab stars 
than what is found in the halo.

While we have done our best to quantify the significance of detected halo
groups, we emphasize that these groups are just candidates whose authenticity
still needs to be verified. This verification can be done by analyzing
metallicities and velocities of RR Lyrae stars obtained from a spectroscopic
followup (e.g., \citealt{ses10b, ses12}). If a spatial group is real, then its
stars should also cluster in the velocity and metallicity space. Such followup
studies are highly encouraged and should provide more conclusive evidence on the
nature of detected groups.

In this work, we used the sample of LINEAR RR Lyrae stars to study the halo
structure and substructures, but its usefulness goes beyond these simple
applications. For example, light curves of LINEAR RRab stars are well sampled
and should allow a robust decomposition into a Fourier series. In turn, the
Fourier components can be used to estimate the metallicity of RRab stars via
the \citet{jk96} method. These metallicities can then be used to study the
metallicity distribution of RR Lyrae stars in the halo. Searches for halo
substructures will also benefit from having metallicities as these represent
an additional dimension for clustering algorithms such as EnLink. As RRab stars
in this sample are brighter than 17 mag, they will be observed by the upcoming
surveys such as the LAMOST Experiment for Galactic Understanding and
Exploration (LEGUE; \citealt{den12}) and GAIA \citep{per12}. These surveys will
provide metallicity, proper motion, and parallax measurements for RR Lyrae
stars up to 30 kpc from the Sun and will enable unprecedented studies of the
structure, formation and the evolution of the Galactic halo.

\acknowledgments

B.S. thanks NSF grant AST-0908139 to Judith G.~Cohen for partial support.
\v{Z}.I. acknowledges support by NSF grants AST-0707901 and AST-1008784 to the
University of Washington, by NSF grant AST-0551161 to LSST for design and
development activity, and by the Croatian National Science Foundation grant
O-1548-2009. A.C.B. acknowledges support from NASA ADP grant NNX09AC77G. The
LINEAR program is funded by the National Aeronautics and Space Administration at
MIT Lincoln Laboratory under Air Force Contract FA8721-05-C-0002. Opinions,
interpretations, conclusions and recommendations are those of the authors and
are not necessarily endorsed by the United States Government.

\bibliographystyle{apj}
\bibliography{ms}

\begin{thebibliography}{66}
\expandafter\ifx\csname natexlab\endcsname\relax\def\natexlab#1{#1}\fi

\bibitem[{{Bla{\v z}ko}(1907)}]{bla07}
{Bla{\v z}ko}, S. 1907, Astronomische Nachrichten, 175, 325

\bibitem[{{Buchler} \& {Koll{\'a}th}(2011)}]{bk11}
{Buchler}, J.~R., \& {Koll{\'a}th}, Z. 2011, \apj, 731, 24

\bibitem[{{Bullock} {et~al.}(2001){Bullock}, {Kravtsov}, \& {Weinberg}}]{bkw01}
{Bullock}, J.~S., {Kravtsov}, A.~V., \& {Weinberg}, D.~H. 2001, \apj, 548, 33

\bibitem[{{Cacciari} {et~al.}(2005){Cacciari}, {Corwin}, \& {Carney}}]{cac05}
{Cacciari}, C., {Corwin}, T.~M., \& {Carney}, B.~W. 2005, \aj, 129, 267

\bibitem[{{Catelan}(2009)}]{cat09}
{Catelan}, M. 2009, \apss, 320, 261

\bibitem[{{Chaboyer}(1999)}]{chaboyer99}
{Chaboyer}, B. 1999, Post-Hipparcos Cosmic Candles, 237, 111

\bibitem[{{Chadid} {et~al.}(2010){Chadid}, {Benk{\H o}}, {Szab{\'o}},
  {Papar{\'o}}, {Chapellier}, {Kolenberg}, {Poretti}, {Bono}, {Le Borgne},
  {Trinquet}, {Artemenko}, {Auvergne}, {Baglin}, {Debosscher}, {Grankin},
  {Guggenberger}, \& {Weiss}}]{cha10}
{Chadid}, M., {Benk{\H o}}, J.~M., {Szab{\'o}}, R., {et~al.} 2010, \aap, 510,
  A39

\bibitem[{{Chun} {et~al.}(2010){Chun}, {Kim}, {Sohn}, {Park}, {Han}, {Kim},
  {Lee}, {Lee}, {Lee}, \& {Sohn}}]{chu10}
{Chun}, S.-H., {Kim}, J.-W., {Sohn}, S.~T., {et~al.} 2010, \aj, 139, 606

\bibitem[{{Clement} {et~al.}(2001){Clement}, {Muzzin}, {Dufton}, {Ponnampalam},
  {Wang}, {Burford}, {Richardson}, {Rosebery}, {Rowe}, \& {Hogg}}]{cle01}
{Clement}, C.~M., {Muzzin}, A., {Dufton}, Q., {et~al.} 2001, \aj, 122, 2587

\bibitem[{{De Lee}(2008)}]{dle08}
{De Lee}, N. 2008, PhD thesis, Michigan State University

\bibitem[{{Deason} {et~al.}(2011){Deason}, {Belokurov}, \& {Evans}}]{dea11}
{Deason}, A.~J., {Belokurov}, V., \& {Evans}, N.~W. 2011, \mnras, 416, 2903

\bibitem[{{Deason} {et~al.}(2012){Deason}, {Belokurov}, {Evans}, \&
  {An}}]{dea12}
{Deason}, A.~J., {Belokurov}, V., {Evans}, N.~W., \& {An}, J. 2012, \mnras,
  424, L44

\bibitem[{{Deng} {et~al.}(2012){Deng}, {Newberg}, {Liu}, {Carlin}, {Beers},
  {Chen}, {Chen}, {Christlieb}, {Grillmair}, {Guhathakurta}, {Han}, {Hou},
  {Lee}, {L{\'e}pine}, {Li}, {Liu}, {Pan}, {Sellwood}, {Wang}, {Wang}, {Yang},
  {Yanny}, {Zhang}, {Zhang}, {Zheng}, \& {Zhu}}]{den12}
{Deng}, L.-C., {Newberg}, H.~J., {Liu}, C., {et~al.} 2012, Research in
  Astronomy and Astrophysics, 12, 735

\bibitem[{{Drake} {et~al.}(2013){Drake}, {Catelan}, {Djorgovski}, {Torrealba},
  {Graham}, {Belokurov}, {Koposov}, {Mahabal}, {Prieto}, {Donalek}, {Williams},
  {Larson}, {Christensen}, \& {Beshore}}]{dra13}
{Drake}, A.~J., {Catelan}, M., {Djorgovski}, S.~G., {et~al.} 2013, \apj, 763,
  32

\bibitem[{{Duffau} {et~al.}(2006){Duffau}, {Zinn}, {Vivas}, {Carraro},
  {M{\'e}ndez}, {Winnick}, \& {Gallart}}]{duf06}
{Duffau}, S., {Zinn}, R., {Vivas}, A.~K., {et~al.} 2006, \apjl, 636, L97

\bibitem[{{Freeman} \& {Bland-Hawthorn}(2002)}]{fbh02}
{Freeman}, K., \& {Bland-Hawthorn}, J. 2002, \araa, 40, 487

\bibitem[{Friedman(1984)}]{fri84}
Friedman, J.~H. 1984, A variable span scatterplot smoother, Tech. rep.,
  Laboratory for Computational Statistics, Stanford University, technical
  Report No. 5

\bibitem[{{Grillmair} \& {Johnson}(2006)}]{gj06}
{Grillmair}, C.~J., \& {Johnson}, R. 2006, \apjl, 639, L17

\bibitem[{{Harding} {et~al.}(2001){Harding}, {Morrison}, {Olszewski},
  {Arabadjis}, {Mateo}, {Dohm-Palmer}, {Freeman}, \& {Norris}}]{har01}
{Harding}, P., {Morrison}, H.~L., {Olszewski}, E.~W., {et~al.} 2001, \aj, 122,
  1397

\bibitem[{{Harris}(1996)}]{har96}
{Harris}, W.~E. 1996, \aj, 112, 1487

\bibitem[{{Helmi}(2008)}]{hel08}
{Helmi}, A. 2008, \aapr, 15, 145

\bibitem[{{Helmi} \& {White}(1999)}]{helmi99}
{Helmi}, A., \& {White}, S.~D.~M. 1999, \mnras, 307, 495

\bibitem[{{Ivezi{\'c}} {et~al.}(2012){Ivezi{\'c}}, {Beers}, \&
  {Juri{\'c}}}]{ibj12}
{Ivezi{\'c}}, {\v Z}., {Beers}, T.~C., \& {Juri{\'c}}, M. 2012, \araa, 50, 251

\bibitem[{{Ivezi\'c} {et~al.}(2013){Ivezi\'c}, {Connoly}, {VanderPlas}, \&
  {Gray}}]{ive13}
{Ivezi\'c}, {\v{Z}}., {Connoly}, A., {VanderPlas}, J., \& {Gray}, A. 2013,
  Statistics, Data Mining, and Machine Learning in Astronomy: A Practical Guide
  for the Analysis of Survey Data, Princeton (Princeton University Press)

\bibitem[{{Ivezi{\'c}} {et~al.}(2005){Ivezi{\'c}}, {Vivas}, {Lupton}, \&
  {Zinn}}]{ive05}
{Ivezi{\'c}}, {\v Z}., {Vivas}, A.~K., {Lupton}, R.~H., \& {Zinn}, R. 2005,
  \aj, 129, 1096

\bibitem[{{Ivezi{\'c}} {et~al.}(2000){Ivezi{\'c}}, {Goldston}, {Finlator},
  {Knapp}, {Yanny}, {McKay}, {Amrose}, {Krisciunas}, {Willman}, {Anderson},
  {Schaber}, {Erb}, {Logan}, {Stubbs}, {Chen}, {Neilsen}, {Uomoto}, {Pier},
  {Fan}, {Gunn}, {Lupton}, {Rockosi}, {Schlegel}, {Strauss}, {Annis},
  {Brinkmann}, {Csabai}, {Doi}, {Fukugita}, {Hennessy}, {Hindsley}, {Margon},
  {Munn}, {Newberg}, {Schneider}, {Smith}, {Szokoly}, {Thakar}, {Vogeley},
  {Waddell}, {Yasuda}, {York}, \& {SDSS Collaboration}}]{ive00}
{Ivezi{\'c}}, {\v Z}., {Goldston}, J., {Finlator}, K., {et~al.} 2000, \aj, 120,
  963

\bibitem[{{Ivezi{\'c}} {et~al.}(2008){Ivezi{\'c}}, {Sesar}, {Juri{\'c}},
  {Bond}, {Dalcanton}, {Rockosi}, {Yanny}, {Newberg}, {Beers}, {Allende
  Prieto}, {Wilhelm}, {Lee}, {Sivarani}, {Norris}, {Bailer-Jones}, {Re
  Fiorentin}, {Schlegel}, {Uomoto}, {Lupton}, {Knapp}, {Gunn}, {Covey},
  {Smith}, {Miknaitis}, {Doi}, {Tanaka}, {Fukugita}, {Kent}, {Finkbeiner},
  {Munn}, {Pier}, {Quinn}, {Hawley}, {Anderson}, {Kiuchi}, {Chen}, {Bushong},
  {Sohi}, {Haggard}, {Kimball}, {Barentine}, {Brewington}, {Harvanek},
  {Kleinman}, {Krzesinski}, {Long}, {Nitta}, {Snedden}, {Lee}, {Harris},
  {Brinkmann}, {Schneider}, \& {York}}]{ive08}
{Ivezi{\'c}}, {\v Z}., {Sesar}, B., {Juri{\'c}}, M., {et~al.} 2008, \apj, 684,
  287

\bibitem[{{Johnston} {et~al.}(2008){Johnston}, {Bullock}, {Sharma}, {Font},
  {Robertson}, \& {Leitner}}]{joh08}
{Johnston}, K.~V., {Bullock}, J.~S., {Sharma}, S., {et~al.} 2008, \apj, 689,
  936

\bibitem[{{Johnston} {et~al.}(1996){Johnston}, {Hernquist}, \& {Bolte}}]{jhb96}
{Johnston}, K.~V., {Hernquist}, L., \& {Bolte}, M. 1996, \apj, 465, 278

\bibitem[{{Jurcsik} \& {Kovacs}(1996)}]{jk96}
{Jurcsik}, J., \& {Kovacs}, G. 1996, \aap, 312, 111

\bibitem[{{Juri{\'c}} {et~al.}(2008){Juri{\'c}}, {Ivezi{\'c}}, {Brooks},
  {Lupton}, {Schlegel}, {Finkbeiner}, {Padmanabhan}, {Bond}, {Sesar},
  {Rockosi}, {Knapp}, {Gunn}, {Sumi}, {Schneider}, {Barentine}, {Brewington},
  {Brinkmann}, {Fukugita}, {Harvanek}, {Kleinman}, {Krzesinski}, {Long},
  {Neilsen}, {Nitta}, {Snedden}, \& {York}}]{jur08}
{Juri{\'c}}, M., {Ivezi{\'c}}, {\v Z}., {Brooks}, A., {et~al.} 2008, \apj, 673,
  864

\bibitem[{{Kafle} {et~al.}(2012){Kafle}, {Sharma}, {Lewis}, \&
  {Bland-Hawthorn}}]{kaf12}
{Kafle}, P.~R., {Sharma}, S., {Lewis}, G.~F., \& {Bland-Hawthorn}, J. 2012,
  \apj, 761, 98

\bibitem[{{Kafle} {et~al.}(2013){Kafle}, {Sharma}, {Lewis}, \&
  {Bland-Hawthorn}}]{kaf13}
---. 2013, \mnras, 430, 2973

\bibitem[{{Keller} {et~al.}(2008){Keller}, {Murphy}, {Prior}, {Da Costa}, \&
  {Schmidt}}]{kel08}
{Keller}, S.~C., {Murphy}, S., {Prior}, S., {Da Costa}, G., \& {Schmidt}, B.
  2008, \apj, 678, 851

\bibitem[{{Kinemuchi} {et~al.}(2006){Kinemuchi}, {Smith}, {Wo{\'z}niak},
  {McKay}, \& {ROTSE Collaboration}}]{kin06}
{Kinemuchi}, K., {Smith}, H.~A., {Wo{\'z}niak}, P.~R., {McKay}, T.~A., \&
  {ROTSE Collaboration}. 2006, \aj, 132, 1202

\bibitem[{{Kolenberg} {et~al.}(2011){Kolenberg}, {Bryson}, {Szab{\'o}},
  {Kurtz}, {Smolec}, {Nemec}, {Guggenberger}, {Moskalik}, {Benk{\H o}},
  {Chadid}, {Jeon}, {Kiss}, {Kopacki}, {Nuspl}, {Still},
  {Christensen-Dalsgaard}, {Kjeldsen}, {Borucki}, {Caldwell}, {Jenkins}, \&
  {Koch}}]{kol11}
{Kolenberg}, K., {Bryson}, S., {Szab{\'o}}, R., {et~al.} 2011, \mnras, 411, 878

\bibitem[{{Kollmeier} {et~al.}(2012){Kollmeier}, {Szczygiel}, {Burns}, {Gould},
  {Thompson}, {Preston}, {Sneden}, {Crane}, {Dong}, {Madore}, {Morrell},
  {Prieto}, {Shectman}, {Simon}, \& {Villanueva}}]{kol12}
{Kollmeier}, J.~A., {Szczygiel}, D.~M., {Burns}, C.~R., {et~al.} 2012, ArXiv
  e-prints (arXiv:1208.2689)

\bibitem[{{Lauchner} {et~al.}(2006){Lauchner}, {Powell}, \& {Wilhelm}}]{lpw06}
{Lauchner}, A., {Powell}, Jr., W.~L., \& {Wilhelm}, R. 2006, \apjl, 651, L33

\bibitem[{{Mayer} {et~al.}(2002){Mayer}, {Moore}, {Quinn}, {Governato}, \&
  {Stadel}}]{may02}
{Mayer}, L., {Moore}, B., {Quinn}, T., {Governato}, F., \& {Stadel}, J. 2002,
  \mnras, 336, 119

\bibitem[{{Miceli} {et~al.}(2008){Miceli}, {Rest}, {Stubbs}, {Hawley}, {Cook},
  {Magnier}, {Krisciunas}, {Bowell}, \& {Koehn}}]{mic08}
{Miceli}, A., {Rest}, A., {Stubbs}, C.~W., {et~al.} 2008, \apj, 678, 865

\bibitem[{{Newberg} {et~al.}(2002){Newberg}, {Yanny}, {Rockosi}, {Grebel},
  {Rix}, {Brinkmann}, {Csabai}, {Hennessy}, {Hindsley}, {Ibata}, {Ivezi{\'c}},
  {Lamb}, {Nash}, {Odenkirchen}, {Rave}, {Schneider}, {Smith}, {Stolte}, \&
  {York}}]{new02}
{Newberg}, H.~J., {Yanny}, B., {Rockosi}, C., {et~al.} 2002, \apj, 569, 245

\bibitem[{{Odenkirchen} {et~al.}(1997){Odenkirchen}, {Brosche}, {Geffert}, \&
  {Tucholke}}]{ode97}
{Odenkirchen}, M., {Brosche}, P., {Geffert}, M., \& {Tucholke}, H.-J. 1997,
  \na, 2, 477

\bibitem[{{Oosterhoff}(1939)}]{oos39}
{Oosterhoff}, P.~T. 1939, The Observatory, 62, 104

\bibitem[{{Perryman}(2002)}]{per12}
{Perryman}, M.~A.~C. 2002, \apss, 280, 1

\bibitem[{Press {et~al.}(1992)Press, Flannery, Teukolsky, \&
  Vetterling}]{pre92}
Press, W., Flannery, B., Teukolsky, S., \& Vetterling, W. 1992, Numerical
  Recipes in C: The Art of Scientific Computing, Numerical Recipes in C: The
  Art of Scientific Computing (Cambridge University Press)

\bibitem[{{Preston} {et~al.}(1991){Preston}, {Shectman}, \& {Beers}}]{psb91}
{Preston}, G.~W., {Shectman}, S.~A., \& {Beers}, T.~C. 1991, \apj, 375, 121

\bibitem[{{Reimann}(1994)}]{rei94}
{Reimann}, J.~D. 1994, PhD thesis, University of California, Berkeley.

\bibitem[{{Schlegel} {et~al.}(1998){Schlegel}, {Finkbeiner}, \&
  {Davis}}]{SFD98}
{Schlegel}, D.~J., {Finkbeiner}, D.~P., \& {Davis}, M. 1998, \apj, 500, 525

\bibitem[{{Sesar} {et~al.}(2011{\natexlab{a}}){Sesar}, {Juri{\'c}}, \&
  {Ivezi{\'c}}}]{sji11}
{Sesar}, B., {Juri{\'c}}, M., \& {Ivezi{\'c}}, {\v Z}. 2011{\natexlab{a}},
  \apj, 731, 4

\bibitem[{{Sesar} {et~al.}(2011{\natexlab{b}}){Sesar}, {Stuart}, {Ivezi{\'c}},
  {Morgan}, {Becker}, \& {Wo{\'z}niak}}]{ses11}
{Sesar}, B., {Stuart}, J.~S., {Ivezi{\'c}}, {\v Z}., {et~al.}
  2011{\natexlab{b}}, \aj, 142, 190

\bibitem[{{Sesar} {et~al.}(2010{\natexlab{a}}){Sesar}, {Vivas}, {Duffau}, \&
  {Ivezi{\'c}}}]{ses10b}
{Sesar}, B., {Vivas}, A.~K., {Duffau}, S., \& {Ivezi{\'c}}, {\v Z}.
  2010{\natexlab{a}}, \apj, 717, 133

\bibitem[{{Sesar} {et~al.}(2007){Sesar}, {Ivezi{\'c}}, {Lupton}, {Juri{\'c}},
  {Gunn}, {Knapp}, {DeLee}, {Smith}, {Miknaitis}, {Lin}, {Tucker}, {Doi},
  {Tanaka}, {Fukugita}, {Holtzman}, {Kent}, {Yanny}, {Schlegel}, {Finkbeiner},
  {Padmanabhan}, {Rockosi}, {Bond}, {Lee}, {Stoughton}, {Jester}, {Harris},
  {Harding}, {Brinkmann}, {Schneider}, {York}, {Richmond}, \& {Vanden
  Berk}}]{ses07}
{Sesar}, B., {Ivezi{\'c}}, {\v Z}., {Lupton}, R.~H., {et~al.} 2007, \aj,, 134,
  2236

\bibitem[{{Sesar} {et~al.}(2010{\natexlab{b}}){Sesar}, {Ivezi{\'c}}, {Grammer},
  {Morgan}, {Becker}, {Juri{\'c}}, {De Lee}, {Annis}, {Beers}, {Fan}, {Lupton},
  {Gunn}, {Knapp}, {Jiang}, {Jester}, {Johnston}, \& {Lampeitl}}]{ses10a}
{Sesar}, B., {Ivezi{\'c}}, {\v Z}., {Grammer}, S.~H., {et~al.}
  2010{\natexlab{b}}, \apj, 708, 717

\bibitem[{{Sesar} {et~al.}(2012){Sesar}, {Cohen}, {Levitan}, {Grillmair},
  {Juri{\'c}}, {Kirby}, {Laher}, {Ofek}, {Surace}, {Kulkarni}, \&
  {Prince}}]{ses12}
{Sesar}, B., {Cohen}, J.~G., {Levitan}, D., {et~al.} 2012, \apj, 755, 134

\bibitem[{{Sharma} \& {Johnston}(2009)}]{sha09}
{Sharma}, S., \& {Johnston}, K.~V. 2009, \apj, 703, 1061

\bibitem[{{Sharma} {et~al.}(2011){Sharma}, {Johnston}, {Majewski}, {Bullock},
  \& {Mu{\~n}oz}}]{sha11}
{Sharma}, S., {Johnston}, K.~V., {Majewski}, S.~R., {Bullock}, J., \&
  {Mu{\~n}oz}, R.~R. 2011, \apj, 728, 106

\bibitem[{{Skrutskie} {et~al.}(2006){Skrutskie}, {Cutri}, {Stiening},
  {Weinberg}, {Schneider}, {Carpenter}, {Beichman}, {Capps}, {Chester},
  {Elias}, {Huchra}, {Liebert}, {Lonsdale}, {Monet}, {Price}, {Seitzer},
  {Jarrett}, {Kirkpatrick}, {Gizis}, {Howard}, {Evans}, {Fowler}, {Fullmer},
  {Hurt}, {Light}, {Kopan}, {Marsh}, {McCallon}, {Tam}, {Van Dyk}, \&
  {Wheelock}}]{skr06}
{Skrutskie}, M.~F., {Cutri}, R.~M., {Stiening}, R., {et~al.} 2006, \aj, 131,
  1163

\bibitem[{Smith(2004)}]{smi04}
Smith, H. 2004, RR Lyrae Stars, Cambridge Astrophysics (Cambridge University
  Press)

\bibitem[{{S{\'o}dor} {et~al.}(2011){S{\'o}dor}, {Jurcsik}, {Szeidl},
  {V{\'a}radi}, {Henden}, {Vida}, {Hurta}, {Posztob{\'a}nyi}, {D{\'e}k{\'a}ny},
  \& {Szing}}]{sod11}
{S{\'o}dor}, {\'A}., {Jurcsik}, J., {Szeidl}, B., {et~al.} 2011, \mnras, 411,
  1585

\bibitem[{{Suntzeff} {et~al.}(1991){Suntzeff}, {Kinman}, \& {Kraft}}]{sun91}
{Suntzeff}, N.~B., {Kinman}, T.~D., \& {Kraft}, R.~P. 1991, \apj, 367, 528

\bibitem[{{S{\"u}veges} {et~al.}(2012){S{\"u}veges}, {Sesar}, {V{\'a}radi},
  {Mowlavi}, {Becker}, {Ivezi{\'c}}, {Beck}, {Nienartowicz}, {Rimoldini},
  {Dubath}, {Bartholdi}, \& {Eyer}}]{suv12}
{S{\"u}veges}, M., {Sesar}, B., {V{\'a}radi}, M., {et~al.} 2012, \mnras, 424,
  2528

\bibitem[{{Szczygie{\l}} {et~al.}(2009){Szczygie{\l}}, {Pojma{\'n}ski}, \&
  {Pilecki}}]{spp09}
{Szczygie{\l}}, D.~M., {Pojma{\'n}ski}, G., \& {Pilecki}, B. 2009, \actaa, 59,
  137

\bibitem[{{Vivas} \& {Zinn}(2006)}]{vz06}
{Vivas}, A.~K., \& {Zinn}, R. 2006, \aj, 132, 714

\bibitem[{{Vivas} {et~al.}(2001){Vivas}, {Zinn}, {Andrews}, {Bailyn}, {Baltay},
  {Coppi}, {Ellman}, {Girard}, {Rabinowitz}, {Schaefer}, {Shin}, {Snyder},
  {Sofia}, {van Altena}, {Abad}, {Bongiovanni}, {Brice{\~n}o}, {Bruzual},
  {Della Prugna}, {Herrera}, {Magris}, {Mateu}, {Pacheco}, {S{\'a}nchez},
  {S{\'a}nchez}, {Schenner}, {Stock}, {Vicente}, {Vieira}, {Ferr{\'{\i}}n},
  {Hernandez}, {Gebhard}, {Honeycutt}, {Mufson}, {Musser}, \&
  {Rengstorf}}]{viv01}
{Vivas}, A.~K., {Zinn}, R., {Andrews}, P., {et~al.} 2001, \apjl, 554, L33

\bibitem[{{Watkins} {et~al.}(2009){Watkins}, {Evans}, {Belokurov}, {Smith},
  {Hewett}, {Bramich}, {Gilmore}, {Irwin}, {Vidrih}, {Wyrzykowski}, \&
  {Zucker}}]{wat09}
{Watkins}, L.~L., {Evans}, N.~W., {Belokurov}, V., {et~al.} 2009, \mnras, 398,
  1757

\bibitem[{{Wu} {et~al.}(2002){Wu}, {Wang}, \& {Chen}}]{wwc02}
{Wu}, Z.-Y., {Wang}, J.-J., \& {Chen}, L. 2002, \cjaa, 2, 216

\end{thebibliography}

\clearpage

\begin{deluxetable}{lrrcrrrrrrrrcc}
\rotate
\setlength{\tabcolsep}{0.02in} 
\tabletypesize{\scriptsize}
\tablecolumns{14}
\tablewidth{0pc}
\tablecaption{Positions and light curve parameters of LINEAR RR Lyrae stars\label{table1}}
\tablehead{
\colhead{LINEAR objectID$^a$} &
\colhead{R.A.$^b$} & \colhead{Dec$^b$} &
\colhead{Type} &
\colhead{Period} & \colhead{HJD$_0^c$} &
\colhead{Amplitude$^d$} & \colhead{$m_0^e$} &
\colhead{Template ID$^f$} &
\colhead{rExt$^g$} &
\colhead{$\langle m\rangle^h$} &
\colhead{Distance$^i$} &
\colhead{Oosterhoff class$^j$} &
\colhead{Group ID$^k$} \\
\colhead{} & \colhead{(deg)} & \colhead{(deg)} & \colhead{} & \colhead{(day)} &
\colhead{(day)} & \colhead{(mag)} & \colhead{(mag)} & \colhead{} &
\colhead{(mag)} & \colhead{(mag)} & \colhead{(kpc)} & \colhead{} & \colhead{}}
\startdata
29848 & 119.526418 & 46.962232 & ab & 0.557021 & 53802.775188 & 0.598 & 16.020 & 4068023 & 0.145 & 16.129 & 12.76 & 1 & 0 \\
32086 & 119.324018 & 47.095636 & ab & 0.569258 & 53774.779119 & 0.721 & 14.535 & 32086   & 0.186 & 14.705 & 6.62  & 1 & 0
\enddata
\tablenotetext{a}{Identification number referencing this object in the LINEAR database. ObjectIDs of RR Lyrae stars taken from the LINEAR Catalog of Variable
Stars (Palaversa, L.~et al., submitted to AJ) end with a ``*'' symbol.}
\tablenotetext{b}{Equatorial J2000.0 right ascension and declination.}
\tablenotetext{c}{Reduced heliocentric Julian date of maximum brightness (HJD$_0$ - 2400000).}
\tablenotetext{d}{Amplitude measured from the best-fit LINEAR template.}
\tablenotetext{e}{Maximum brightness measured from the best-fit LINEAR template (not corrected for interstellar medium extinction).}
\tablenotetext{f}{Best-fit LINEAR template ID number.}
\tablenotetext{g}{Extinction in the SDSS $r$ band calculated using the \citet{SFD98} dust map.}
\tablenotetext{h}{Flux-averaged magnitude (corrected for interstellar medium extinction as $\langle m\rangle = \langle m\rangle_{not\, corrected} - {\rm rExt}$.}
\tablenotetext{i}{Heliocentric distance (see Section~\ref{distances}).}
\tablenotetext{j}{Oosterhoff type (1 or 2 for RRab stars, 0 for RRc stars).}
\tablenotetext{k}{Substructure group ID (0 for stars not associated with a substructure).}
\tablecomments{Table~\ref{table1} is published in its entirety in the electronic edition of the Journal. A portion is shown here for guidance regarding its form and content.}
\end{deluxetable}

\clearpage

\begin{deluxetable}{lcccccccc}
\setlength{\tabcolsep}{0.02in} 
\tabletypesize{\scriptsize}
\tablecolumns{9}
\tablewidth{0pc}
\tablecaption{Halo substructures detected in LINEAR\label{table2}}
\tablehead{
\colhead{Group} &
\colhead{R.A.$^a$} & \colhead{Dec$^a$} & \colhead{Distance$^a$} &
\colhead{Radius$^b$} &
\colhead{${\rm N_{stars}}$} & \colhead{Significance$^c$} &
\colhead{Near} & \colhead{$N_{OoII}/N_{OoI}^d$} \\
\colhead{} & \colhead{(deg)} & \colhead{(deg)} & \colhead{(kpc)} &
\colhead{(kpc)} & \colhead{} & \colhead{} & \colhead{} & \colhead{}
}
\startdata
1 & 252.434435 & 39.649328 & 10.1 & 1.3 & 35 & 2.58 & ? & 0.52 \\
2 & 240.211410 & 16.229577 &  7.7 & 1.0 & 27 & 2.63 & ? & 0.80 \\
3 & 198.052741 & 20.254224 & 15.7 & 1.1 & 25 & 1.55 & NGC 5053 or NGC 5024 (M53) & 0.79 \\
4 & 205.408486 & 28.409365 & 10.4 & 1.0 & 21 & 1.77 & NGC 5272 (M3) & 0.05 \\
5 & 252.816666 & 30.449727 &  6.8 & 0.7 & 20 & 1.88 & NGC 6205 (M13) & 0.33 \\
6 & 185.795386 & -0.282617 & 15.4 & 1.4 & 16 & 1.60 & Virgo Stellar Stream & 0.23 \\
7 & 149.429078 & 54.102182 &  6.0 & 1.5 & 10 & 1.42 & ? & 0.25
\enddata
\tablenotetext{a}{Right ascension, declination and the heliocentric distance of
the peak in number density, where the number density has been measured by
EnLink.}
\tablenotetext{b}{Radius of the group, estimated as the median distance of stars
from the peak in number density.}
\tablenotetext{c}{Significance of the group, as measured by EnLink.}
\tablenotetext{d}{Ratio of Oosterhoff type II to Oosterhoff type I RRab stars.
For the full LINEAR RRab sample, this ratio is 0.25.}
\end{deluxetable}

\clearpage


\clearpage

\clearpage

\clearpage

\clearpage

\clearpage

\clearpage

\clearpage

\clearpage

\clearpage

\clearpage

\clearpage

\clearpage

\clearpage

\clearpage

\clearpage

\clearpage

\clearpage

\end{document}